\documentclass[twocolumn]{aastex62}

\newcommand{\BG}[1]{{#1}}
\newcommand{\EuS}[1]{{#1}}

\newcommand{\norm}[1]{\|#1\|}
\newcommand{\Gaia}{{\em Gaia}}
\graphicspath{{./}{figures/}}

\received{July 2, 2018}
\revised{\today}
\accepted{\today}
\submitjournal{ApJ}

\shorttitle{A large moving group within the Lower Centaurus Crux association}
\shortauthors{Goldman et al.}

\begin{document}

\title{{A large moving group within the Lower Centaurus Crux association}}

\correspondingauthor{Bertrand Goldman}
\email{goldman@mpia.de}

\author[0000-0002-2729-7276]{Bertrand Goldman}
\affiliation{Max-Planck-Institut f\"ur Astronomie, K\"onigstuhl 17, 69117 Heidelberg, Germany}
\affiliation{Observatoire astronomique de Strasbourg, Universit\'e de Strasbourg - CNRS UMR 7550, 11 rue de l'Universit\'e, 67000, Strasbourg, France}
\nocollaboration

\author{Siegfried R\"oser}
\affiliation{Zentrum f\"ur Astronomie der Universit\"at Heidelberg, Landessternwarte, K\"{o}nigstuhl 12, 69117 Heidelberg, Germany}
\affiliation{Zentrum f\"ur Astronomie der Universit\"at Heidelberg, Astronomisches Rechen-Institut, M\"{o}nchhofstra\ss{}e 12-14, 69120 Heidelberg, Germany}
\nocollaboration

\author{Elena Schilbach}
\affiliation{Zentrum f\"ur Astronomie der Universit\"at Heidelberg, Landessternwarte, K\"{o}nigstuhl 12, 69117 Heidelberg, Germany}
\affiliation{Zentrum f\"ur Astronomie der Universit\"at Heidelberg, Astronomisches Rechen-Institut, M\"{o}nchhofstra\ss{}e 12-14, 69120 Heidelberg, Germany}
\nocollaboration

\author{Attila C. Mo\'or}
\affiliation{Konkoly Observatory, Research Centre for Astronomy and Earth Sciences, Hungarian Academy of Sciences, H-1121 Budapest, Konkoly Thege Mikl\'os \'ut 15-17, Hungary}
\nocollaboration

\author[0000-0002-1493-300X]{Thomas Henning}
\affiliation{Max-Planck-Institut f\"ur Astronomie, K\"onigstuhl 17, 69117 Heidelberg, Germany}

\begin{abstract}

Scorpius-Centaurus is the nearest OB association and its hundreds of members were divided into sub-groups, including Lower Centaurus Crux.
Here we {study the dynamics of the Lower Centaurus Crux area}. We report the \EuS{revelation of a large moving group containing {more than} 1800} intermediate- and low-mass young stellar objects and brown dwarfs, that escaped identification until \Gaia\ DR2 allowed to perform a kinematic and photometric selection.
{We investigate the stellar and substellar content of this moving group using the \Gaia\ DR2 astrometric and {photometric} measurements. }
{The median distance of the members is 114.5~pc and 80\% lie between 102 and 135~pc from the Sun.} Our new members cover a mass range of 5\,M$_\odot$ to 0.02\,M$_\odot$, and add up to a total mass of about 700\,M$_\odot$.
The present-day mass function follows a log-normal law with m$_c$ = 0.22 {M$_\odot$} and $\sigma$ = 0.64. We find more than 200 brown dwarfs in our sample. The star formation rate had its maximum of $8\times10^{-5}\rm M_\odot yr^{-1}$ at about 9\,Myr ago,
We grouped the new members in four denser subgroups, which 
have increasing age from 7 to 10\,Myr, surrounded by ``free-floating" young stars with mixed ages. 
\BG{Our isochronal ages, now based on accurate parallaxes, are compatible with several earlier studies of the region.}
The whole complex is presently expanding, and the expansion started between 8 to 10\,Myr ago. Two hundred members show infrared excess compatible with circumstellar disks from full to debris disks. This discovery provides a large sample of nearby young stellar and sub-stellar objects for disk and exoplanet studies. 

\end{abstract}


\section{Introduction} \label{sec:intro}

Co-eval structures of the Solar neighborhood are useful entities for various fields in astronomy. Either gravitationally bound (a.k.a. open clusters) or not (moving groups), they provide samples of stars that share the same composition, motion, and age.
Possible deviations from this set of parameters also give insights into the formation of these structures: the velocity dispersion may allow to constrain the age and location of the birth place of the population; the composition and age dispersion may reveal a more complicated formation and multiple stages of star formation. 
Co-evolution on the other hand lets us constrain the stellar evolutionary models.
To find such structures in the Solar neighborhood allows spectroscopic follow-up of high spectral  resolution and/or on small telescopes, 
and easy detection of low-mass members, 
making them particularly valuable.

For almost a century, the Scorpius-Centaurus OB-association (Sco-Cen) has been
known as an ensemble of young, massive early-type stars, and which should hold the keys of star formation. \citet{1964ARA&A...2..213B} divided the huge nearby association into three subgroups, Upper Scorpius (US), Upper Centaurus Lupus (UCL) and Lower Centaurus Crux (LCC). Being on the southern hemisphere the quality of the astrometric data historically was always poorer than for associations on the northern hemisphere. Only with the advent of the Hipparcos data the situation changed, but only for a subset of the brightest stars. In their fundamental paper, using the Hipparcos measurements, \citet{1999AJ....117..354D} presented an overview of the OB-associations in the wider neighborhood of the Sun. For Scorpius-Centaurus they obtained mean distances of 145 pc for US, 140 pc for UCL and  118 pc for LCC. \citet{2016MNRAS.461..794P} determined  median ages of 11 (US), 16 (UCL), and 17\,Myr (LCC), though each has a considerable
spread of ages. Other authors obtained  ages as low as 5\,Myr for US
\citep{Preib99,Preib08}, and 16 (UCL) to\,18 Myr (LCC) \citep{Sarto03, Mamaj02}.

In their introduction \citet{2018MNRAS.476..381W} give an excellent overview on the importance of the Sco-Cen complex for questions related to the modes of star formation. Using the observations from \Gaia\,DR1 they confirmed that all three subgroups US, UCL and LCC are gravitationally unbound, and found neither evidence for expansion of the subgroups nor that they, or part of them, were formed by the disruption of star clusters.

\EuS{
Using the Hipparcos data and a convergent point method, \citet{1999AJ....117..354D} 
found extended moving groups in the area of Sco-Cen, one in each sub-group.
However, they and subsequent authors could only detect stars with masses typically above
0.7 M$_\odot$. The situation was best described by \citet{Preib08}:
``the surface of the census of low-mass Sco-Cen members has barely been scratched."
In this paper, we use the results of  \Gaia\,DR2 \citep{Brown18} to uncover a large population of young stellar objects located in the LCC area of the Sco-Cen region. About 70 per cent of these stars belong to a large moving group and assemble in several subgroups of increasing age.} 

The paper is structured as follows: 
In Section\,\ref{detect} we describe the identification of the \EuS{ moving group and its sub-groups} in the TGAS and \Gaia\ DR2 catalogs; in Section\,\ref{massage} we determine and discuss the mass functions and age distributions of the groups.
In Section\,\ref{members} we discuss some interesting members, in particular those surrounded by circumstellar disks, and binaries. In Section\,\ref{motion} we study the internal dynamics and expansion of the groups. {In Section \ref{conclusion} we summarize and discuss the major results of our paper. }

\section{Detection of the Crux star forming region} \label{detect}
\subsection{Astrometric detection}\label{astdet}
Parallel to the cluster search program described in \citet{2016A&A...595A..22R}, we ran an alternative search process better adapted to the finding of nearby co-moving groups. We give only a rough description of the procedure here as it is not crucial for this paper. The idea behind was to select a grid of
convergent points homogeneously distributed over the sphere and to search for stars whose
positions, proper motions and parallaxes are consistent with a given convergent point. We then selected stars which
have at least 5 other stars within a radius of 10 pc around them and whose tangential velocities are consistent with the same convergent point. These we regarded as seeds of possible
co-moving groups. Our search was restricted to stars having a parallax larger than 5.5 mas in TGAS. On the whole sky we found 706 of these seeds. Here we concentrate onto the area of the Scorpius-Centaurus OB-association (Sco-Cen) between Galactic longitude 280\degr and 360\degr and Galactic latitude from $-20$\degr  to + 30\degr.

\begin{figure*}[htb!]
  \centering
   \plotone{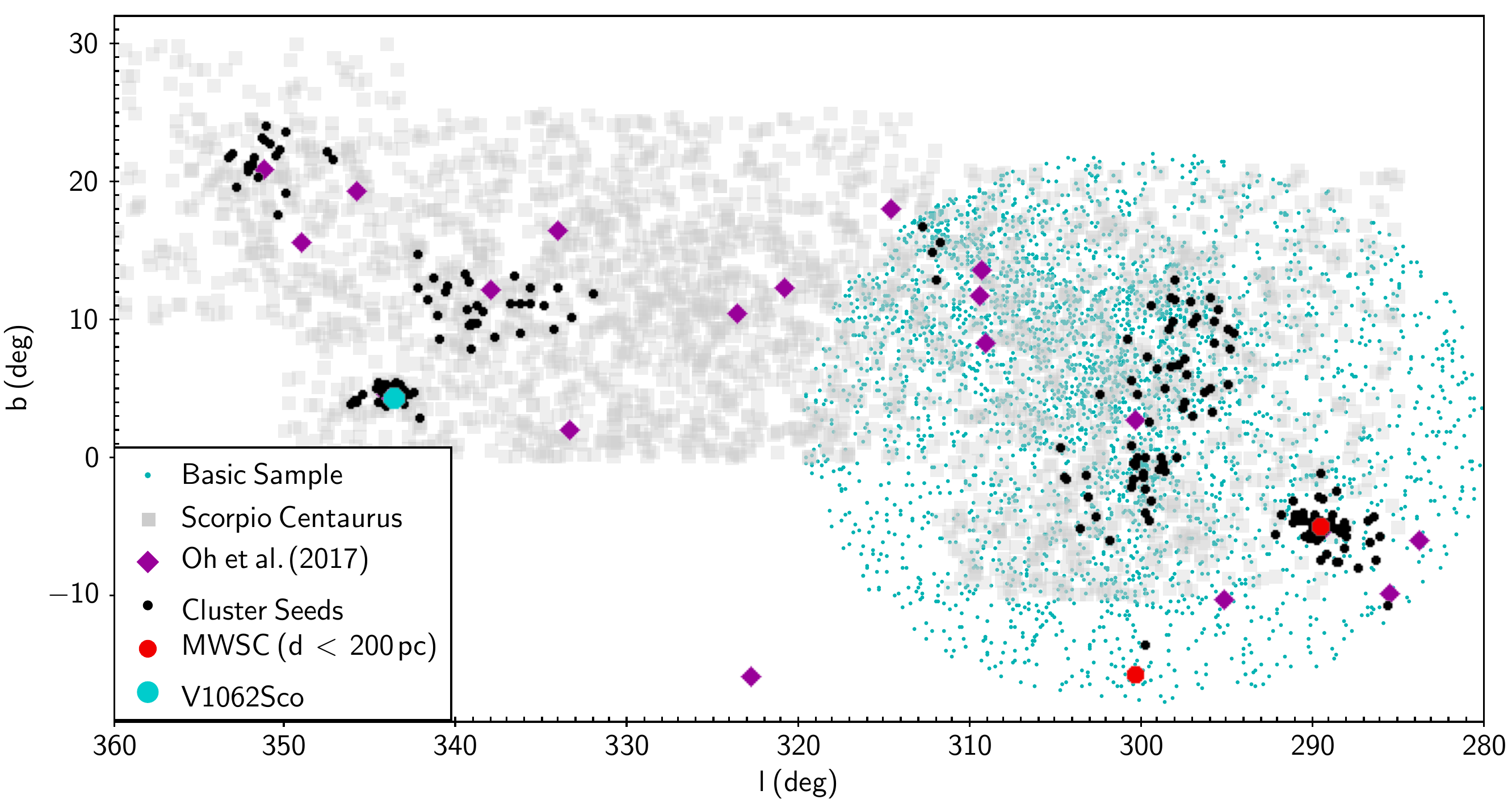}
      \caption{The Scorpius-Centaurus Region. The axes are Galactic longitude and Galactic latitude in degrees. The grey squares in the background show the stars from \citet{Hooge00} and outline the Sco-Cen association. The black dots are the seeds of our cluster search in TGAS (see text). The small cyan dots outline the region studied in this paper and show the stars of our Basic Sample (BS, see text). The large cyan dot at $(l,b) = (344\deg,+4\deg)$ is the young compact moving group around V1062Sco \citep{2018A&A...614A..81R}. The large violet diamonds are the groups found by \citet{2017AJ....153..257O}, and the large red dots mark open clusters in MWSC, with distances less than 200 pc. }
 \label{SCOCEN}
\end{figure*}
   
In Fig.~\ref{SCOCEN} we present the result of our search in the area of Sco-Cen. In the background we show the stars from \citet{Hooge00} (US, UCL, LCC) by large gray squares. This serves as an indication of the extend of the Sco-Cen association.  The seeds found by our search in TGAS are shown as black dots. Isolated seeds stand for small co-moving groups with not more than 6 stars in a 10-pc radius around the seed in TGAS. Spatial concentrations of seeds may suggest larger coherent groups. For instance, the concentration marked by the large cyan dot 
at $(l,b) = (344\deg,+4\deg)$ is the young compact moving group around V1062Sco \citep{2018A&A...614A..81R} at a distance of 175 pc.
For comparison we also plot the centers of {the two} open clusters contained in MWSC \EuS{\citep[Milky Way Star Clusters,][]{2013A&A...558A..53K}}
up to 200~pc from the Sun (large red dots), as well as the groups found by \citet{2017AJ....153..257O} with more than 5 neighbors (violet diamonds). In parts, their and our findings are overlapping.
We retrieved the open cluster IC 2391 in MWSC at Galactic longitude $l = 289\deg$. Even the poorly populated cluster Feigelson~1 \citep[also known as the $\epsilon$ Chameleontis group,][]{2013ApJS..209...26F} at $(l,b)=(300\deg,-15\deg)$ was found as one seed. 
Of the other seeds, the concentrations at $(l,b)=(300\deg,-1\deg)$ and the large group at $(l,b)=(297\deg,+8\deg)$ seemed to be the most interesting, because of their proximity to the Sun; the mean 
of the TGAS parallaxes of these groups were 9.5 and 9.1\,mas, respectively. We called the southern group Group A and the northern one Group B.
The {violet diamond} in between Group A and B corresponds to the center of ``Group 4" in \citet{2017AJ....153..257O}. Their group 4
actually comprises our groups A and B, and in their notation it has a size of 114 (number of stars in the group). It is worth mentioning that already \citet{1997ESASP.402..545C} found a moving group of 33 nearby A-type dwarfs at a mean distance  of 105 pc at $(l,b) = (304\deg,+3\deg)$.
For our Group A we determined the phase space coordinates on the basis of the TGAS positions, proper motions and parallaxes, as well as with radial velocities
retrieved from SIMBAD, and found mean values of:
\EuS{
\small{
\begin{equation}
\begin{array}{rcccl}
\vec{R_c} & = & (\xi_c,\eta_c,\zeta_c) & = & (-47.17 , -5.23, -96.41)\,{\rm pc}, \\
\vec{V_c} & = & (v_{\xi,c},v_{\eta,c},v_{\zeta,c}) & = & ( -2.97, 18.90, -14.76)\,{\rm km\,s^{-1}}
\label{CPcoo}
\end{array} 
\end{equation}}
in the Barycentric equatorial coordinate system with the
$\xi$-axis pointing to the Vernal Equinox, the $\eta$-axis to $\alpha = 90\deg$ 
and the $\zeta$-axis points to the Equatorial North Pole.} The vector $\vec{R_c}$ is measured in parsec and is the position vector of the center of the group, and $\vec{V_c}$, in $\rm km~s^{-1}$, is the vector of the space velocity of the group; {$\vec{V_{c}}/{\norm{\vec{V_{c}}}}$} being the Cartesian equatorial coordinates of the convergent point.
   
\Gaia\ DR2 now enables to study the groups A and B in great detail. In a first step we made a query of the \Gaia\,DR2 catalog consisting of a cone search around $(\alpha,\delta)$ =  (186.5\degr, -60.5\degr) with a radius of 20 degrees. \EuS{The center of the cone has been chosen midway between our groups A and B from TGAS and the radius large enough to contain both groups.} Further we applied the following restrictions: for the trigonometric parallax \mbox{$\varpi \geq 7 $mas}; for the proper motions  -43 mas/y $\leq \mu_{\alpha*} \leq$ -20 mas/y ($\mu_{\alpha*} \equiv \mu_\alpha \cos\delta$),  and -35 $\leq \mu_{\delta} \leq$ +12 mas/y. 
The proper motion and parallax cuts were chosen to be consistent with the distribution of stars identified as members of these probable young moving groups we revealed in TGAS.
\EuS{This query yielded 20,138 objects and we call this set of objects ``Crux Cone".}
For the further processing we followed the procedures 
described in \citet[Chapter 4.3 and Appendix C, Figs C.1 and C.2]{2018arXiv180409366L} to obtain a stellar sample cleaned from possible artifacts.

First, we required the ``visibility periods used" to be equal to or larger than 7 \citep[Eq. (11, ii) in][]{2018arXiv180409366L}, 
which left us with 19,878 objects. 
The ``unit weight error"-cut \citep[cf. Eq C.1 in][]{2018arXiv180409366L}
removed a considerable portion of dubious entries, and 7,324 sources remained. Instead of the factor of 1.2 in this equation,
we used a factor of 1.3 which seems more appropriate to our sample \EuS{and keeps some 40 \BG{additional} stars in the magnitude range $10.0<G<19.5$\,mag.}
{Among these 40 stars, 11 (among 1844) will be selected in our final sample.
It seems therefore likely that the astrometric parameters for those 11 stars found to be both young and comoving (with a narrow selection, as will be shown below) are correct and those stars should be kept in our analysis. The 1\% increase in the statistics has of course no incidence on the conclusion of this paper. }
As a next step we applied the ``flux excess ratio"-cut \citep[Eq.~C.2 in][]{2018arXiv180409366L} which reduced the sample 
to 3946 objects. Finally we discarded three stars with relative errors of parallaxes \mbox{$\sigma_\varpi/\varpi$} larger than 10 per cent. 
In summary, from the originally 20,138 entries in our search cone 3,943 sources survived these filterings,  forming an 
astrometrically clean sample. For the subsequent discussion we always refer to this sample as ``the Basic Sample (BS)". {The stars of the BS are shown in Fig.~\ref{SCOCEN} as small cyan dots and outline the area of our search.}

\subsection{Candidate selection}\label{candsel}
\EuS{
In Fig.~\ref{HRDnew_fig2.pdf} we show the Color Magnitude Diagram (CMD) of the 3,943 stars of the BS, where we plot the absolute magnitude $M_{\rm G}$ (using the parallaxes from DR2) versus the
color $G-G_{\rm RP}$. 
For colors redder than about 0.4 mag two separate sequences show up, which we attribute to young
stars (upper sequence) and old stars (lower sequence). We strengthen this in  Fig.~\ref{HRDnew_fig2.pdf}, where we also plot the theoretical isochrones of CIFIST \citep{2015A&A...577A..42B} and Parsec \citep{2017ApJ...835...77M}.
The red lines are the CIFIST isochrones for 10 Myr and 400 Myr, while the black lines are the 10 Myr and 400 Myr Parsec isochrones. From a comparison between the two sets of isochrones for different ages, we found that the brighter part of the CMD is appropriately described by the Parsec isochrones,
whereas the CIFIST isochrones better fit the pre-main-sequence part for stars less massive than 1.4\,M$_{\odot}$.
 {Here we note that the CIFIST isochrones use the pre-launch filter profiles of \citet{Jordi10}, while the Parsec isochrones use the profiles of \citet{Evans18}, both in the Vega system.}

The difference between the two observed sequences is too large to be explained by unresolved binaries of main sequence stars in this field. Consequently, the upper sequences must be attributed to young pre-main sequence stars. To separate the old and young population we made a cut along the
40 Myr CIFIST isochrone from the faintest stars up to the locus ( $G-G_{\rm RP}$, $M_{\rm G}$) = (0.34, 3.22), corresponding to 1.4 M$_{\odot}$, and continue on the Parsec 40 Myr isochrone for brighter stars. The actual separation is shown in Fig.~\ref{HRDnew_fig3.pdf}. 
We additionally discarded a small number of scattered stars which we attribute to older red giants. 
The sub-sample of young stars (2,659) in the BS is called ``Young", the complement (1,284 stars) ``Old". We note that beginning at the turnover at ( $G-G_{\rm RP}$, $M_{\rm G}$) = (0.75,6.0) there may be increasing contamination by old binaries to the sample ``Young". Moreover, a small number of main sequence stars with
$G-G_{\rm RP} \leq $ 0.34 may be also field stars, which can be identified and excluded by their different kinematics.
\begin{figure}[htb!]
 \centering
 \plotone{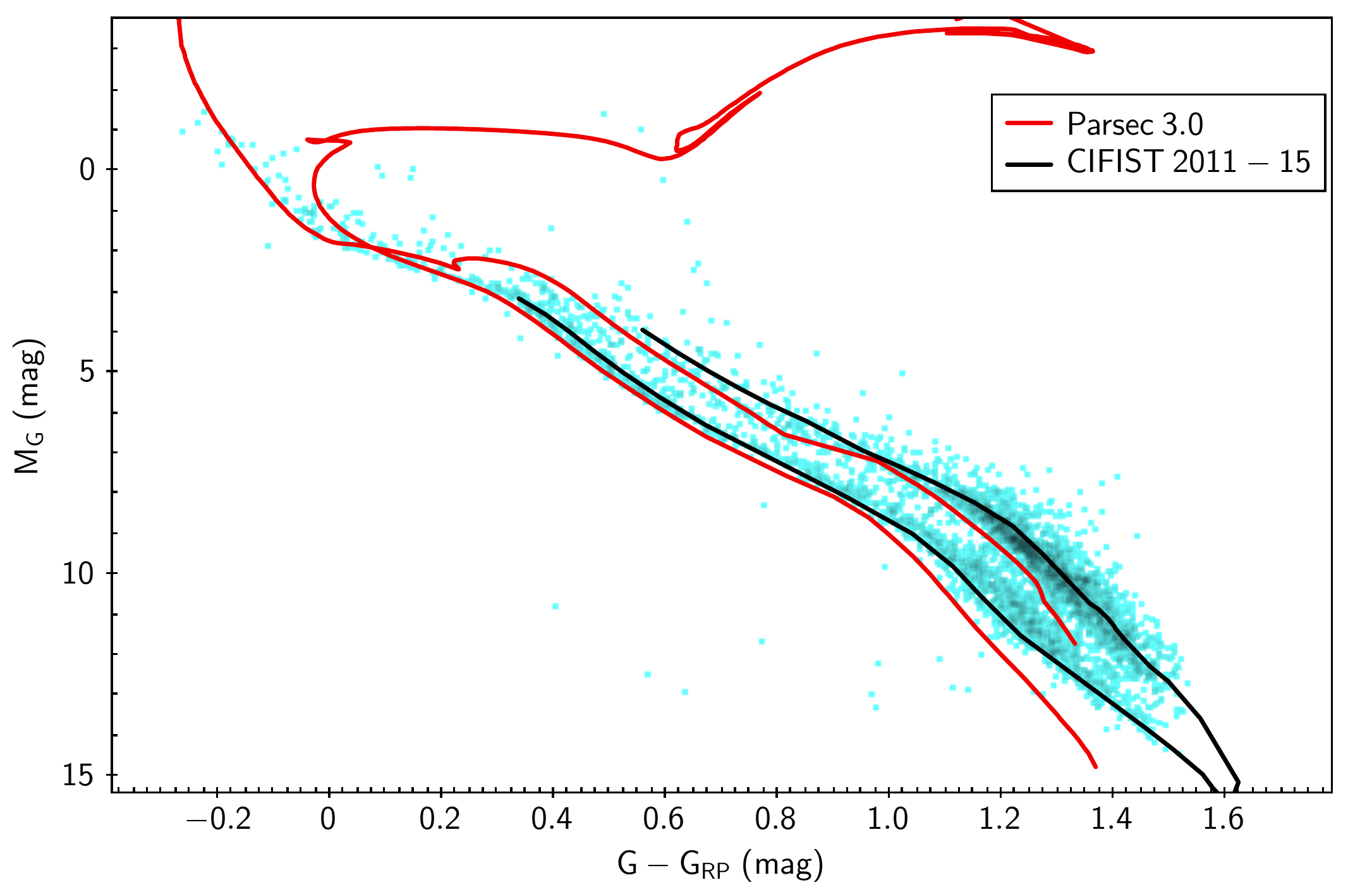}
 \caption{Absolute magnitudes M$_G$ vs. G-G$_{RP}$ of the 3,943 stars of the basic sample (BS). The black curves are the CIFIST isochrones from \citet{2015A&A...577A..42B} for 10 Myr (upper curve) and 400 Myr (lower curve). The red lines are the 10 Myr (upper curve) and 400 Myr (lower curve) isochrones from Parsec3.0 \citep{2017ApJ...835...77M}.}
         \label{HRDnew_fig2.pdf}
\end{figure}  
\begin{figure}[htb!]
 \centering
 \plotone{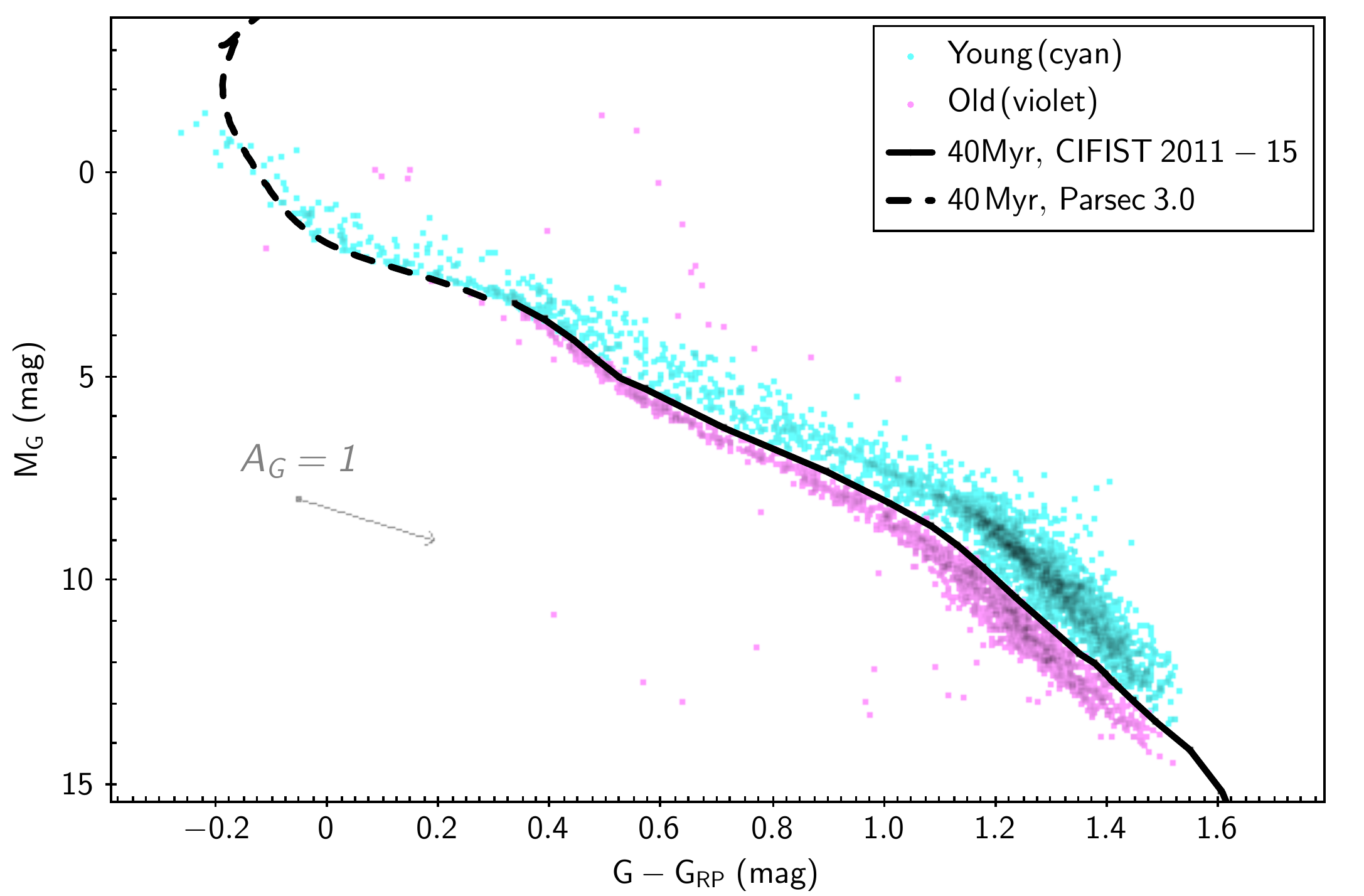}
      \caption{Absolute magnitudes M$_G$ vs. G-G$_{RP}$ of the 3,943 stars of the basic sample (BS). The 40 Myr isochrone of \citep[][solid line, masses less than 1.4 M$_\odot$]{2015A&A...577A..42B}  and Parsec 3.0 (dashed line, masses higher than 1.4 M$_\odot$) separates the young population (cyan) from the old population (violet). An A$_G$ = 1 mag reddening vector is shown by an arrow.}
         \label{HRDnew_fig3.pdf}
   \end{figure}    
We discussed the CMDs without considering a possible extinction towards the sample
of our stars from the BS.
One may expect that its influence is rather small since 
our stars are all nearby ($\varpi \geq 7 $mas). 
On the other hand, they are actually
located within the Sco-Cen association and
may not be far away from or behind gas and molecular clouds. 
However, from Figure~\ref{HRDnew_fig2.pdf} we note a good agreement between the 400 Myr isochrone and the main sequence of older BS 
stars of intermediate masses, and there is no
reason to assume a significant extinction towards the stars in the BS. To verify this result, we also
considered CMDs in other pass-bands including the near-infrared (J, K$_S$) from 2MASS \citep{2006AJ....131.1163S},  and Gaia BP.
Nevertheless, any attempt to introduce a correction
for de-reddening led to a poorer fittings of isochrones to observations in the CMDs.
Of course we cannot exclude that some individual stars may be significantly reddened, especially if we consider the corresponding reddening vector in Figure~\ref{HRDnew_fig3.pdf}.   {The extinction data published in \Gaia\,DR2 does not allow} to check {this option} as \citet{2018arXiv180409374A} mention in their paper.
To summarize, we refrained from de-reddening the stars in the BS.}

\subsubsection{Kinematic analysis}\label{kinematics}
\EuS{
To reveal the members of the potential moving group characterized by Eq.~\ref{CPcoo}, we then studied the kinematics of the stars in the BS. Unfortunately, only about 12\% of these stars have accurate radial velocities in \Gaia\,DR2, with mean errors smaller than 2~km~s$^{-1}$. Therefore, in order to identify co-moving stars, we have to rely on criteria based on their tangential velocities {only}. The tangential velocity for each star~$i$ has the components $\kappa\,\mu_{\alpha*,i}/\varpi_i$ and $\kappa\,\mu_{\delta,i}/\varpi_i$ with $\kappa=4.74047$ being the transformation factor from 1~mas~yr$^{-1}$ at 1~kpc to 1~km~s$^{-1}$.
Stars with a common space motion should have tangential velocities directed towards a convergent point, in our case defined by Eq.~\ref{CPcoo}.
We computed the components of the observed tangential velocity parallel and perpendicular to the local vector in the direction of this convergent point by a simple rotation around an angle $\psi$ which we determine below.

We followed the formalism of the convergent point method as described, e.g. in \citet{2011A&A...531A..92R} and transformed
the cartesian velocity vector of the group motion $\vec{V_c}$ from Eq.\,\ref{CPcoo} by 
\begin{equation}
\left[\begin{array}{l}V_{c,\alpha,i} \\ 
V_{c,\delta,i} \\\end{array}\right] = 
\left[\begin{array}{rrr} -\sin\alpha_i & \cos\alpha_i & 0 \\ 
-\cos\alpha_i\sin\delta_i & -\sin\alpha_i\sin\delta_i & \cos\delta_i \\
\end{array}\right]
\,\cdot\,\vec{V_c}.
\label{equ:spacVel3}
\end{equation}
With Eq. (8) from \citet{2009A&A...497..209V} the angle $\psi$ is given as $\psi=\arctan(V_{c,\delta,i}/V_{c,\alpha,i})$, and the
desired rotation reads
\begin{equation}
\left[\begin{array}{l}v_{\parallel,i}\\
 v_{\bot,i}\\\end{array}\right] = 
\left[\begin{array}{rr} \cos\psi & \sin\psi \\
-\sin\psi & \cos\psi \\\end{array}\right] 
\,\cdot\,\left[\begin{array}{l}
\kappa\,\mu_{\alpha*,i}/\varpi_i \\
\kappa\,\mu_{\delta,i}/\varpi_i \\\end{array}\right].
\label{equ:rotpsi}
\end{equation}

In an ideal case, where all the stars of a group would exactly move
along the vector $\vec{V_{c}}$, $v_\bot$ would become zero. 
This implies that we a priori know the vector $\vec{V_{c}}$ with highest accuracy. But even in this case, internal velocity dispersion and 
uncertainties of the measurements will cause scatter in $v_\bot$.
}

\begin{figure}[htb!]
 \centering
 \plotone{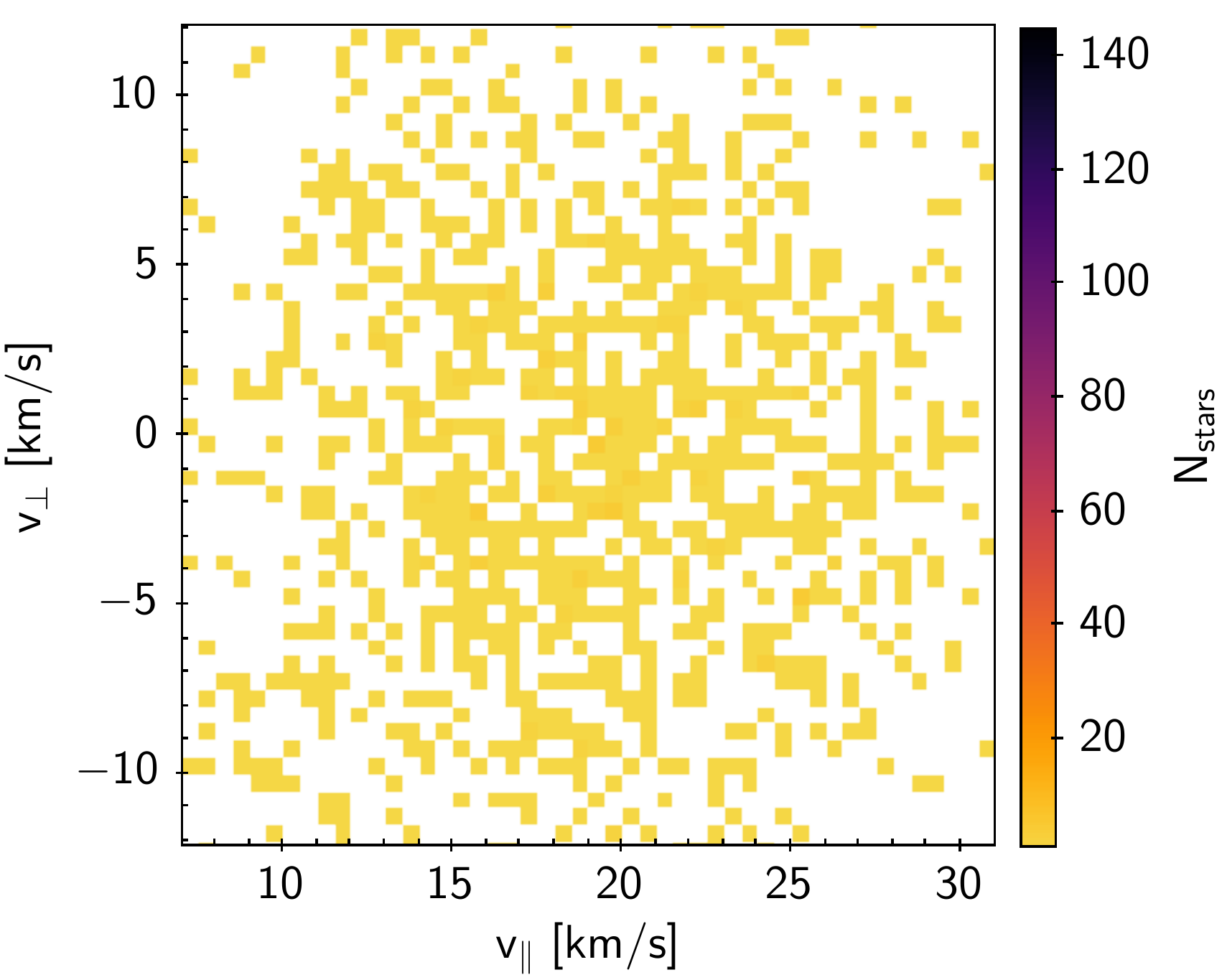}
 \caption{Distribution of the stars from the Old sample in
the $v_\parallel$, $v_\bot$-plane with a binning of $0.5 \times 0.5$ $(\rm km~s^{-1})^2$. $N_{\rm stars}$ is the number of stars per bin. }
 \label{V_par_sen_old}
\end{figure}

\begin{figure}[htb!]
 \centering
 \plotone{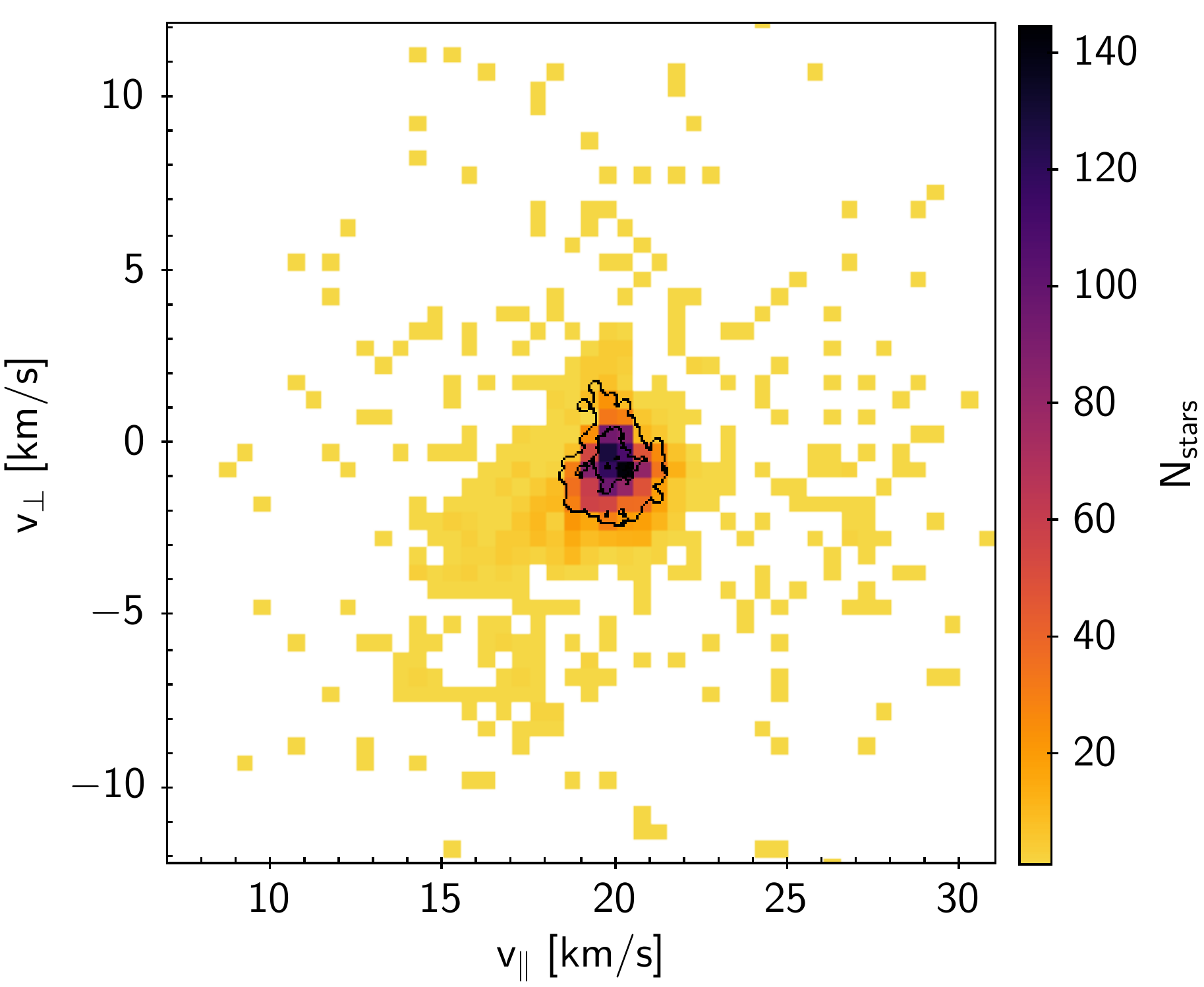}
 \caption{Distribution of the stars from the Young sample in
the $v_\parallel$, $v_\bot$-plane with a binning of $0.5 \times 0.5$ $(km~s^{-1})^2$. N$_{stars}$ is the number of stars per bin.}
         \label{V_par_sen_young}
   \end{figure}

We also determined the covariance matrix for the velocities $v_\parallel$ and  $v_\bot$ according to error propagation from
the covariance matrix of the $\mu_{\alpha*},\, \mu_{\delta}$,\,$\varpi$. From the distribution of the variances we determined the mean and the 90\% percentile of  $\sigma_{v_\parallel}$ as \mbox{$0.25\,\rm km~s^{-1}$} and \mbox{$0.36\,\rm km~s^{-1}$}, respectively. The corresponding values for {$\sigma_{v_\bot}$}  are $0.08\,\rm km~s^{-1}$ and $0.11\,\rm km~s^{-1}$.

Figures~\ref{V_par_sen_old} and \ref{V_par_sen_young} show the distribution of the stars in the Old and the Young samples in
the $v_\parallel$, $v_\bot$-plane with a binning of $0.5 \times 0.5 \,\rm (km~s^{-1})^2$. This bin size is large compared to the variances in $v_\parallel$ and $v_\bot$, as we just showed. So, the distributions in Figures~\ref{V_par_sen_old} and \ref{V_par_sen_young} present the true physical velocity dispersions of the BS. The different behavior of the old and young populations
is obvious. The old stars (left image) are homogeneously distributed on the plane with a typical density of less than 3 stars
per $0.5 \times 0.5 \rm (km~s^{-1})^2$. Such a flat distribution is fully consistent with the broad velocity dispersion of
Galactic field stars. On the other hand, the velocity distribution of the young stars shows an impressive overdensity with a peak at \mbox{$(20.25, -0.75)\,\rm km~s^{-1}$} with a maximum of more than 150 stars per $0.5 \times 0.5 \rm (km~s^{-1})^2$.  
This over-density is a strong signature for an autonomous coherent group consisting essentially of stars younger than about 40 Myr (Fig.~\ref{HRDnew_fig3.pdf}).

In Fig.~\ref{V_par_sen_young} we define this group by manually cutting the stars falling into the peak mentioned above. Indeed they are populating the bins at a level of more than 30 N$_{Stars}$. This cut is indicated by contour lines and the selected sample contains 1844 objects. We call this sample the Crux Moving Group (CMG) henceforth.

\begin{figure}[htb!]
 \centering
 \plotone{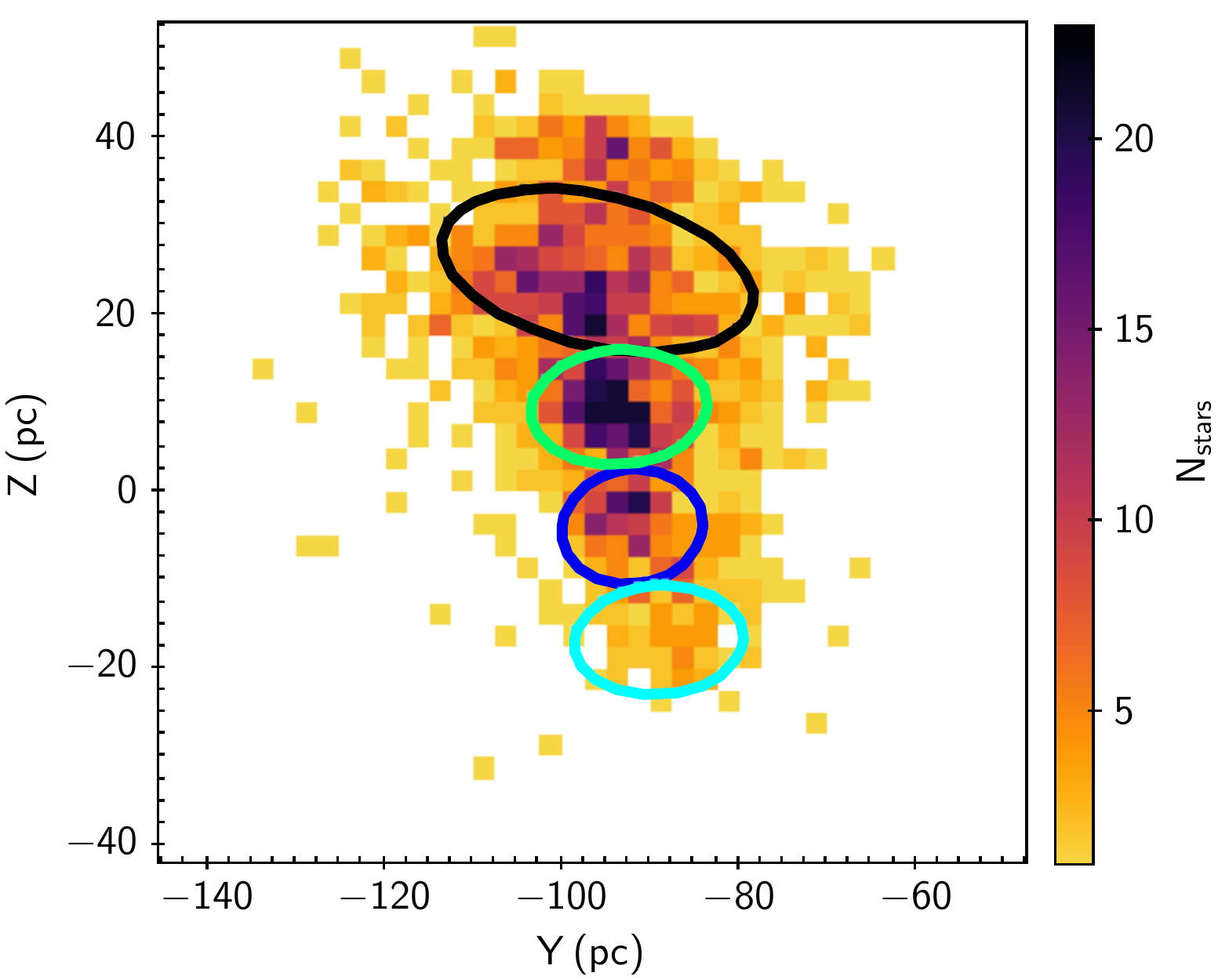}
 \caption{Density distribution of the stars from the CMG projected on to the $Y,Z$ plane (Galactic coordinates). The bin size corresponds to $2.5 \times 2.5$ pc$^2$. N$_{stars}$ is the number of stars per bin. \BG{ The cyan, blue, green, and black lines show the borders of the sub-groups A0, A, B, and C, respectively.}}
 \label{YZ_Vcut_clumps}
\end{figure} 

In the following discussion we use barycentric Galactic Cartesian coordinates $X, Y, Z$ instead of the equatorial coordinates as before. The axes $X, Y, Z$ are directed to the Galactic center, the direction of Galactic rotation and to the Galactic north pole, respectively. The corresponding velocity coordinates are $U, V, W$.
When we compare both samples, we find that, also in space, stars from the Old sample are
distributed homogeneously, whereas the
young stars, and especially the CMG stars, exhibit stronger local concentrations. In
Fig.~\ref{YZ_Vcut_clumps} we present the distribution of the 1844~stars of the
CMG sample \BG{in the YZ plane}. Although these stars were selected in a narrow velocity range, they show a relatively broad distribution in space with several remarkable local over-densities.
This suggested to sub-divide CMG into four subgroups (CMG~A0, CMG~A, CMG~B and CMG~C). \EuS{In Fig.~\ref{YZ_Vcut_clumps} we draw the borders of the four sub-groups with cyan, blue, green, and black lines respectively.} The set of stars from CMG not contained in Groups A0 to C is called CMG Intergroup (CMG~Z). Let us mention here that this subdivision is somehow arbitrary, and was guided by the appearance of these groups already in Fig.~\ref{SCOCEN}.

{
The excellent quality \EuS{($\langle\varpi/\sigma_\varpi\rangle \geq $10)} of the individual parallaxes from \Gaia\ DR2 in the region of CMG allows to invert them into individual distances. The four sub-groups are not only distinguished by their position on the sky, but also by their different distances from the Sun. Figure~\ref{dist_histo} shows the distribution of the individual distances of the stars in the four sub-groups. There is a clear trend in distance from A0 as closest from the Sun to the largest sub-group C as farthest. Altogether, the whole CMG, including the Intergroup members, extends from 102 to 135 pc from the sun (10 and 90\% percentiles of the parallax distribution). The sub-groups A0 to C are more compact. 
   \begin{figure}[htb!]
   \centering
   \plotone{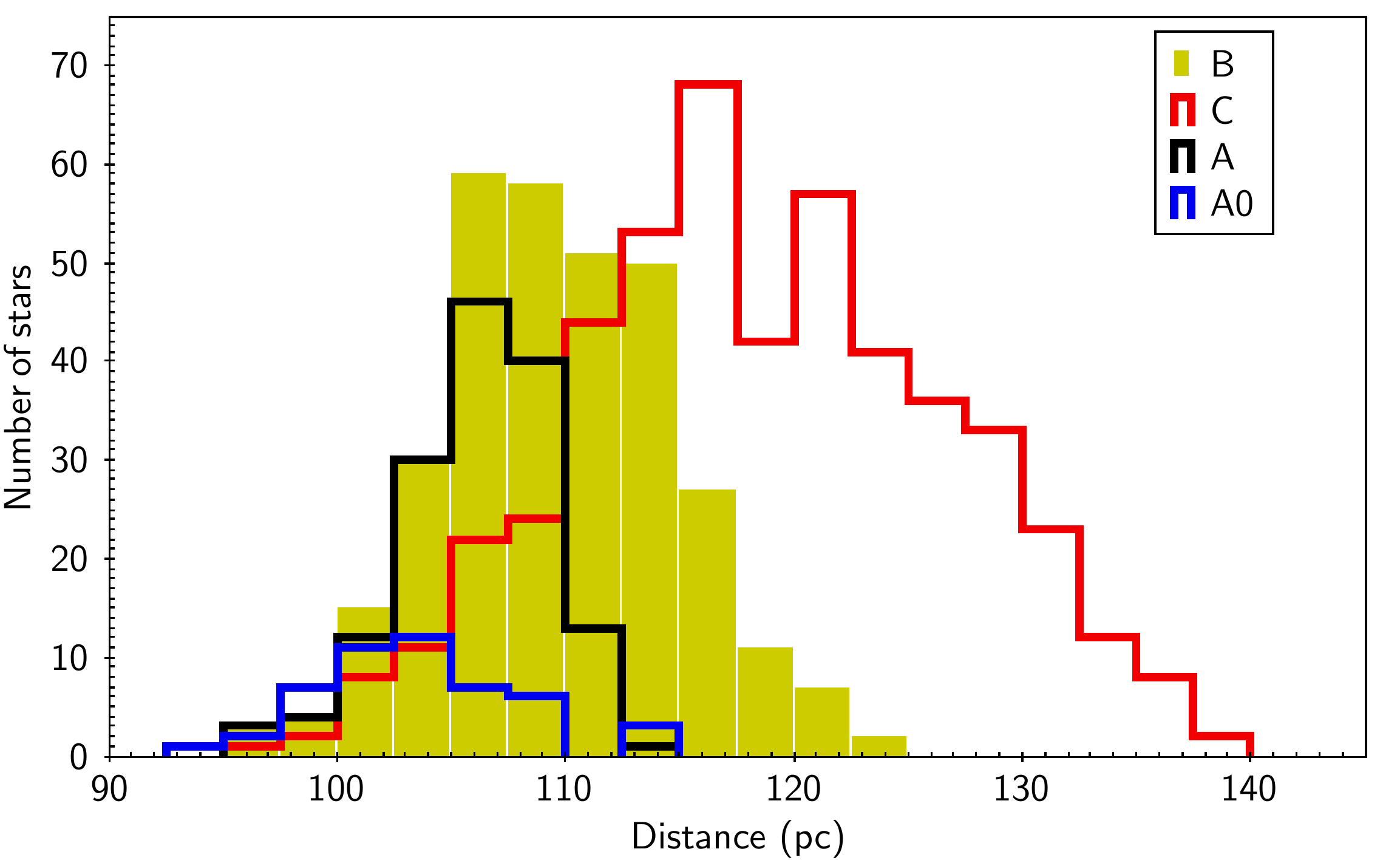}
      \caption{Distribution of the distances of the individual stars in the sub-groups of the CMG derived from the trigonometric parallaxes from \Gaia\ DR2.}
         \label{dist_histo}
   \end{figure} }
The characteristic data for these groups are given in Table\,\ref{TabGroups}, while the data for individual members available in electronic format are described in Table\,\ref{memtable}. The spatial coordinates of the centers of the four subgroups are the means of the individual $X,Y,Z$ of the $N_{\rm Stars}$ in each subgroup.
The space velocities $U,V,W$ of the groups are the averages of the individual velocities of those stars that have reliable radial velocity measurements {in \Gaia\ DR2}. \EuS{In Table~\ref{TabGroups} we give the standard deviations for the velocities, not the mean errors of the mean velocities. These are much more accurate and range from 0.06 to 0.50 $\rm km\,s^{-1}$ with a median of 0.2 $\rm km\,s^{-1}$.}
\begin{table*}[htb!]
\centering
\caption{Characteristic data  of the sub groups of CMG. 
The \BG{mean} values are followed by their standard deviations.}
\begin{tabular}{c r r@{$\pm$}l r@{$\pm$}l r@{$\pm$}l  r@{$\pm$}l r@{$\pm$}l r@{$\pm$}l c c c}
\hline\hline
  \multicolumn{1}{c}{Subgroup} &
  \multicolumn{1}{c}{N$_{Stars}$} &
  \multicolumn{2}{c}{$\bar{X}$} &
  \multicolumn{2}{c}{$\bar{Y}$} &
  \multicolumn{2}{c}{$\bar{Z}$} &
  \multicolumn{2}{c}{$\bar{U}$} &
  \multicolumn{2}{c}{$\bar{V}$} &
  \multicolumn{2}{c}{$\bar{W}$} &
  \multicolumn{1}{c}{$D_{\rm median}$} &
  \multicolumn{1}{c}{age} &
  \multicolumn{1}{c}{Mass} \\
   &
   &
  \multicolumn{2}{c}{(pc)} &
  \multicolumn{2}{c}{(pc)} &
  \multicolumn{2}{c}{(pc)} &
  \multicolumn{2}{c}{($\rm km\,s^{-1}$)} &
  \multicolumn{2}{c}{($\rm km\,s^{-1}$)} &
  \multicolumn{2}{c}{($\rm km\,s^{-1}$)} &
  \multicolumn{1}{c}{(pc)} &
  \multicolumn{1}{c}{(yr)} &
  \multicolumn{1}{c}{($\rm M_\odot$)} \\ 
\hline
  A0 & 49  & 52 &  3 & -88 & 5 & -17 & 3 & -10.2 & 0.5 & -19.4 & 1.0 & -8.9 & 0.3 & 103.0  &   7.0 &   13  \\ 
  A & 149  & 53 &  3 & -92 & 3 &  -3 & 3 &  -9.5 & 0.7 & -20.1 & 1.1 & -7.9 & 0.3 & 106.5  &   7.2 &   50 \\ 
  B & 317  & 55 &  8 & -94 & 4 & +10 & 3 &  -8.8 & 1.1 & -20.2 & 1.8 & -7.1 & 0.3 & 109.5  &   9.3 &   98  \\ 
  C & 487  & 63 & 11 & -97 & 8 & +23 & 4 &  -8.2 & 1.1 & -20.9 & 1.5 & -6.1 & 0.6 & 118.0  &  10.0 &  249  \\ 
\hline\end{tabular}
\label{TabGroups}
\end{table*}

There are no sharp, physically justified boundaries in space between CMG~A0, A, B and C and the stars from  CMG~Z. However, we will show in the following that, on average, they slightly differ not only by their location and kinematics, but also by their average age.
\EuS{
\subsection{CMG compared with other LCC samples}\label{comp}
Before comparing our results with the work of other authors, we first summarize the selection process that led us to the large moving group CMG. 
The first selection from \Gaia\ DR2 was
a cone search around $(\alpha,\delta)$ =  (186.5\degr, -60.5\degr) with a radius of 20 degrees, trigonometric parallax \mbox{$\varpi \geq 7 $mas}, and proper motions  -43 mas/y $\leq \mu_{\alpha*} \leq$ -20 mas/y ($\mu_{\alpha*} \equiv \mu_\alpha \cos\delta$),  and -35 $\leq \mu_{\delta} \leq$ +12 mas/y (see Sec.~\ref{astdet}). This resulted in the Crux Cone  with 20138 objects. Then we applied quality filters
as recommended by \citet[][also described in Sec.~\ref{astdet}]{2018arXiv180409366L} that led to the Basic sample (BS). These and the other relevant sub-samples
are summarized in Table~\ref{downselect}.
\begin{table}[htb!]
\caption{The selection window for the Crux Group of Young Stars (CMG)}\label{downselect}
\centering
\begin{tabular}{lr}
\hline
  \multicolumn{1}{c}{Dataset} &
  \multicolumn{1}{c}{N\_stars} \\
\hline
\hline
  Crux Cone & 20138\\
  Basic sample (BS) & 3943\\
  ``Old'' & 1284\\
  ``Young'' & 2659\\
  CMG & 1844\\
\hline\end{tabular}
\end{table}
 
The area studied in this paper covers most of LCC and a small part of UCL as shown in Fig.~\ref{SCOCEN}. In the following we compare the stellar content of our samples with the findings of other authors from the literature. A summary of these comparisons is given in Table~\ref{others}. 
\begin{table*}
\centering
\caption{Comparison of the CMG with other datasets of members in LCC. For each publication, we report the {total number of objects of the publication in column~2, and in the following columns the number of those stars included in various subsamples, as defined in the text.} }
\label{others}
\begin{tabular}{lrrrrrrr}
\hline
\hline
   \multicolumn{1}{l}{Publication} &
   \multicolumn{1}{r}{total} &
   \multicolumn{6}{c}{{number of stars in}} \\
&  \multicolumn{1}{r}{number} &
   \multicolumn{1}{l}{area} &
   \multicolumn{1}{r}{Crux Cone} &
   \multicolumn{1}{r}{BS} &
   \multicolumn{1}{r}{old sample} &
   \multicolumn{1}{r}{young sample} &
   \multicolumn{1}{r}{CMG} \\
\hline
   \citet{1999AJ....117..354D}--LCC &  180 & 179 & 129& 123 & 13 & 110 & 72  \\
   \citet{1999AJ....117..354D}--UCL &  221 &  19 & 15 &  15 &  1 &  14 & 10  \\
   \citet{Hooge00}--LCC             & 1036 & 1016& 254& 247 & 48 & 199 & 120 \\
   \citet{Hooge00}--UCL             & 1498 & 174 & 35 &  33 &  4 &  29 & 16  \\
   \citet{2011MNRAS.416.3108R}      &  436 & 156 & 123& 119 &  8 & 111 & 76  \\
   \citet{Mamaj02}                  &  110 &  50 & 41 &  41 &  2 &  39 & 30  \\
   \citet{2012ApJ...746..154P}      &  138 &  68 & 50 &  48 &  6 &  42 & 28  \\
   \citet{2012AJ....144....8S}      &  104 &  67 & 57 &  53 &  0 &  53 & 48  \\
   \citet{2016MNRAS.461..794P}      &  493 & 158 & 120& 118 &  2 & 116 & 95  \\
   \citet{Preib08}--LCC             &  121 & 119 & 93 &  89 &  8 &  81 & 61  \\
   \citet{Preib08}--UCL             &   90 &  5  &  2 &  1  &  0 &   1 & 1   \\
\hline\end{tabular}
\end{table*}
The first column {lists} the publications with which we compare {our sample}. The second column gives the number of stars in the respective dataset. The third column is the {size of} subset of stars from column 2, which are {located} in the cone around $(\alpha,\delta) =  (186.5\degr, -60.5\degr$; radius 20~degrees).
In the fourth column, we list the number of stars from column~3 which have DR2 measurements and fall into the Crux Cone. 
The following columns list the cross-matches of the objects from column 4 with
our BS (column 5), the 'Old' sample (column 6),  the 'Young' sample (column 7), and the CMG (column 8).

In their influential paper \citet{1999AJ....117..354D} applied a convergent point method to the Hipparcos measurements to find moving groups in 12 nearby OB associations, one of them was found in Lower Centaurus Crux. Their LCC moving group contains 180 members. Table~\ref{others} shows that all but one star are in our area, but only 129 are in the Crux Cone. This means that 50 of the de Zeeuw stars have Gaia DR2 parallaxes smaller than 7 mas and/or proper motions outside our selection window. The other six stars did not survive the quality cuts (\citet{2018arXiv180409366L}, see also Section~\ref{astdet}) and, therefore, did not appear in the BS: they are probably too bright ($V<6$\, mag) to get accurate parameters in DR2. However, the majority of the de Zeeuw stars in the Crux Cone (85\%) is classified as young by us, and 2/3 of the latter fulfill the strict kinematic conditions of being members in CMG. About the same holds for the de Zeeuw UCL stars which are in our area. We note here that the faintest star from de Zeeuw in CMG is \object{HIP~108016} (F6/7V, $V=9.78$\,mag). This does not mean that the de Zeeuw sample is anyway complete down to $V\approx 10.0$\,mag. The Hipparcos catalogue is based on an input list with a restriction of 2.5~stars per square degree, a magnitude limit at about 12th magnitude in $V$, and has a completeness limit as bright as $V = 7.3$ mag. 

\citet{Hooge00} made use of the `Astrographic Catalogue + Tycho'
reference catalogue (ACT) and the Tycho Reference Catalogue (TRC), which are complete
to V = 10.5 mag. These catalogues only contain proper motions of a typical accuracy of 3 mas/y, but no parallaxes. So, it is not surprising that among the 1190 Hoogerwerf stars in our area only 289 are in the Crux Cone. About 80\% of them are 'Young' and 136 stars fulfill the strong kinematic restrictions of CMG.

\citet{2011MNRAS.416.3108R} took the data from the Hipparcos catalogue and combined these with radial velocities taken from the 2nd Catalogue of Radial Velocities with Astrometric Data \citep{2007AN....328..889K}. Using this data set and a Bayesian membership
selection method they found 436 members in the whole Scorpius-Centaurus complex. In CMG we find 76 stars out of their 436, 71 of them are common to \citet{1999AJ....117..354D}.
Based upon kinematically and X-ray preselected candidates, \citet{Mamaj02} carried out spectroscopic observations and characterized 110 pre-main-sequence stars as UCL and LCC members with 41 of them in the Crux Cone. All but two stars are classified as young by us also, and 30 of them are members in CMG.
\citet{2012ApJ...746..154P} investigated the 138 F-type members of the \citet{1999AJ....117..354D} sample. Out of the 48 of their stars in our BS, 42 are young and 28 are in CMG.

\citet{2012AJ....144....8S}  
spectroscopically identified 104\,\mbox{G-,} \mbox{K-,} and M-type members of the Scorpius-Centaurus complex based on X-ray and kinematically pre-selected samples. They confirmed the youth of these stars by comparing Li $\lambda$6708 absorption line strengths against those of stars in the TW Hydrae association and the $\beta$ Pictoris moving group. From their 57 members in the Crux Cone, 53 are in our BS, all of them are young, and 48 of them are members in CMG. 

\citet{2016MNRAS.461..794P} undertook a spectroscopic survey to search for new K- and M-type members of Scorpius-Centaurus. They obtained a sample of 493 solar-mass ($\approx$ 0.7-1.3 M$_\odot$) stars, out of which 156 were previously unknown. Of their 493 members 120 are in the Crux Cone, 116 are young and 95 are members in our CMG.

Finally we mention the excellent review of \citet{Preib08} which, in great detail, describes the pre-\Gaia\ DR2 situation with respect to the Scorpius-Centaurus OB association. From their paper, we have extracted the 121 stars they discuss in LCC and the 90 from UCL. Ninety percent of their stars have masses larger than 0.8 M$_\odot$. Out of the 211 stars, 124 are in our area, 95 fall in the Crux Cone, 90 are in the BS, 82 are young and 62 are members in our CMG. 

Among the stars in all these samples which were found in the Crux Cone, the vast majority is also classified as young by us. The confirmation  is especially important when we compare with such LCC samples where members were selected by not only using kinematic criteria (i.e., the last six lines in Table~\ref{others}. Among these stars in the Crux Cone, 96\% are in our basic sample and more than 90\% are in the Young sample. Moreover, about 75\% of these stars are members of the moving group CMG.
The high-mass members found before the advent of \Gaia\ DR2 form only the tip of the iceberg in the CMG moving group. On the other hand, we have another about 800 stars in our Young sample which are not as strictly co-moving as the stars in CMG. They may have a similar star formation history, but may have got high relative velocities compared to the bulk of CMG.} 

\begin{table*}[htb!]
\centering
\caption{Content of the electronic table of the member parameters. Several \Gaia\ DR2 columns, which are included for convenience without any change, are not described below. 
The content of the last seven columns is described in Section\,\ref{disks}. 
\BG{In particular, the VOSA columns report the SED fit using either the CIFIST grid, or for the A- and B-type stars, the ATLAS9 grid.}
Reference for disk identification: 
         1 -- \citet{carpenter2005}, 2 -- \citet{chen2005}, 3 -- \citet{chen2011}, 4 -- \citet{chen2012},
         5 -- \citet{clarke2005}, 6 -- \citet{cotten2016}, 7 -- \citet{cruz-saenz2014}, 8 -- \citet{fujiwara2009},  
         9 -- \citet{gregorio-hetem1992}, 10 -- \citet{mannings1998}, 11 -- \citet{marton2016},
         12 -- \citet{melis2012}, 13 -- \citet{melis2013}, 14 -- \citet{oudmaijer1992},
         15 -- \citet{2016MNRAS.461..794P}, 16 -- \citet{rizzuto2012}, 17 -- \citet{schneider2012}, 18 -- this work 
}
\begin{tabular}{l l c r@{$\:\sim\:$}l c}
\hline
\hline
  \multicolumn{1}{c}{Name} &
  \multicolumn{1}{c}{Description} &
  \multicolumn{1}{c}{Unit} &
  \multicolumn{2}{c}{Range}  \\ 
\hline
 designation & \Gaia\ DR2 designation        & \nodata & \multicolumn{2}{c}{\nodata}  \\
 Group       & group membership            & \nodata & \multicolumn{2}{c}{A0,A,B,C,Z}  \\
 ra, dec,... & \Gaia\ DR2 parameters & \nodata & \multicolumn{2}{c}{\nodata}  \\
 vr\_pred     & predicted radial velocity        &  $\rm km\,s^{-1}$ & 5.56 & 16.6  \\
vpar & observed tangential velocity toward the convergent point & $\rm km\,s^{-1}$ & 18.4 & 21.4 \\
vperp       & \shortstack[b]{observed tangential velocity \\perpendicular to the convergent point direction}    & $\rm km\,s^{-1}$ & $-2.3$ & $+2.0$ \\
e\_vpar     & uncertainty on vpar & $\rm km\,s^{-1}$ & 0.053 & 1.28 \\
e\_vperp    & uncertainty on vperp & $\rm km\,s^{-1}$ & 0.012 & 0.41 \\
vparvperpcor      & (dimensionless) correlation between  & \nodata & $-0.83$ & 0.79 \\
 X\_gal           & X galactic coordinate & pc & +26.1 & +102.7 \\
 Y\_gal           & Y galactic coordinate & pc &  $-134.0$ & $-64.6$ \\
 Z\_gal           & Z galactic coordinate & pc & $-30.6$ & $+51.0$ \\
age\_CIFIST       & age according to the CIFIST isochrones & Gyr & $10^{-3}$ & $10^{-1}$ \\
mass\_CIFIST      & mass according to the CIFIST isochrones & $\rm M_\odot$ & 0.026 & 1.4 \\
Teff\_CIFIST       & effective temperature according to the CIFIST isochrones & K & 2632 & 6766 \\
logL\_CIFIST      & $\log_{10}$ of bolometric luminosity according to the CIFIST isochrones & CGS & $-2.8$ & 0.6 \\
logg\_CIFIST      & $\log_{10}$ of gravity according to the CIFIST isochrones & CGS & 3.4 & 4.9 \\
age\_P            & age according to the Parsec isochrones & yr & $5\,10^{6}$ & $10^{8}$ \\
mass\_P           & (initial) mass according to the Parsec isochrones & $\rm M_\odot$ & 0.1 & 5.3 \\
logL\_P           & $\log_{10}$ of bolometric luminosity according to the Parsec isochrones & $\rm L_\odot$ & $-2.7$ & 3.0\\
logTe\_P          & $\log_{10}$ of effective temperature according to the Parsec isochrones & K & 3.4 & 4.2 \\
logg\_P           & $\log_{10}$ of gravity according to the Parsec isochrones & CGS & 3.8 & 4.8 \\
recovery\_rate    & fraction of recovered stars of same $G$ magnitude & \nodata & 0.648 & 1. \\
age               & best age  & Gyr & $10^{-3}$ & $10^{-1}$ \\
mass              & best mass & $\rm M_\odot$ & 0.026 & 5.3 \\
Teff              & best effective temperature  & K & 2632 & $16.10^3$ \\
LogL              & best effective $\log_{10}$ of bolometric luminosity  & K & $-2.8$ & 3.0 \\
AllWISE\_ID      & ID in the AllWISE catalog & \nodata & \multicolumn{2}{c}{\nodata}  \\
logL\_SED        & $\log_{10} $ bolometric luminosity according to the VOSA fit & $\rm Lbol_\odot$ & $-3.6$ & 2.4 \\
Teff\_SED        & effective temperature according to the VOSA fit & K & 1200 & 7000 \\
disk\_type       & Circumstellar disk classification & \nodata & \multicolumn{2}{c}{\nodata} \\
SIMBAD\_Name     & Primary SIMBAD name & \nodata & \multicolumn{2}{c}{\nodata}\\
Disk\_ref        & Reference for circumstellar disk & \nodata & 1 & 18 \\
\hline\end{tabular}
\label{memtable}
\end{table*}

\section{Stellar masses and ages}  \label{massage}

\subsection{Individual mass and age determination}\label{massdet}

The high photometric quality of \Gaia\ DR2 enables a straightforward fit between the observed absolute magnitudes $G$, $G_{\rm BP}$ and $G_{\rm RP}$ and the ones determined from stellar evolution theory. 
\BG{This method has some limitations: it is of course model dependent and the results will be as good as the assumptions made to derive the isochrones. The past accretion history max influence the results, although \citet{Baraf12} suggests that this may not affect clusters aged 10\,Myr or older, except for the lowest-mass stars \citep{Baraf10}. The impact of variability will be limited by the fact that \Gaia\ is multi-epoch and we used the mean of several measurements. Finally, although a fraction of the binaries are resolved by \Gaia\ (see Section\,\ref{ResBins}), unresolved binaries may appear brighter and younger than if they were resolved. We do not attempt to quantify those effects. }

We used two grids of models: the CIFIST~2011\_2015\footnote{\dataset[phoenix.ens-lyon.fr/Grids/BT-Settl/CIFIST2011\_2015/\\ISOCHRONES/]{https://phoenix.ens-lyon.fr/Grids/BT-Settl/CIFIST2011_2015/ISOCHRONES/}} models for intermediate-mass stars down to the brown dwarf regime \citep{2015A&A...577A..42B}, and the PADOVA tracks {Parsec 3.0}\footnote{\dataset[stev.oapd.inaf.it/cmd]{http://stev.oapd.inaf.it/cmd}} for the higher-mass stars \citep{2017ApJ...835...77M}. We adopted 0.7\,M$_\odot$ as the limit between the two regimes, as the derived mass is identical for both models for that mass. We simultaneously determined the mass and age of each object by searching for the nearest model magnitudes to the data, after having linearly refined the model grid. We used the magnitudes in the filters $G$, $G_{\rm BP}$ and $G_{\rm RP}$ and only considered models up to 100\,Myr. 

We compared the model predictions in all three \Gaia\ bands G, G$_{BP}$ and G$_{RP}$ and made a simply 3-d match between observed and theoretical data with a inverse relative weight of 1, 15. and 2.5 respectively, roughly corresponding to the \Gaia\ photometric uncertainties. 

\BG{We discuss our results and compare with previous age determinations in Section\,\ref{Ages}.}

\subsection{Detection efficiency} \label{deteff}

In order to derive the mass functions of the groups, we first needed to determine our detection efficiency, in particular what fraction of members are lost due the selection cuts we applied  {to obtain a kinematically clean sample}, as a function of the $G$ magnitude and the sky location, which conjunctly for \Gaia\ affects the most the precision and accuracy of the measurements. 

\begin{figure}[htb!]
   \centering
   \plotone{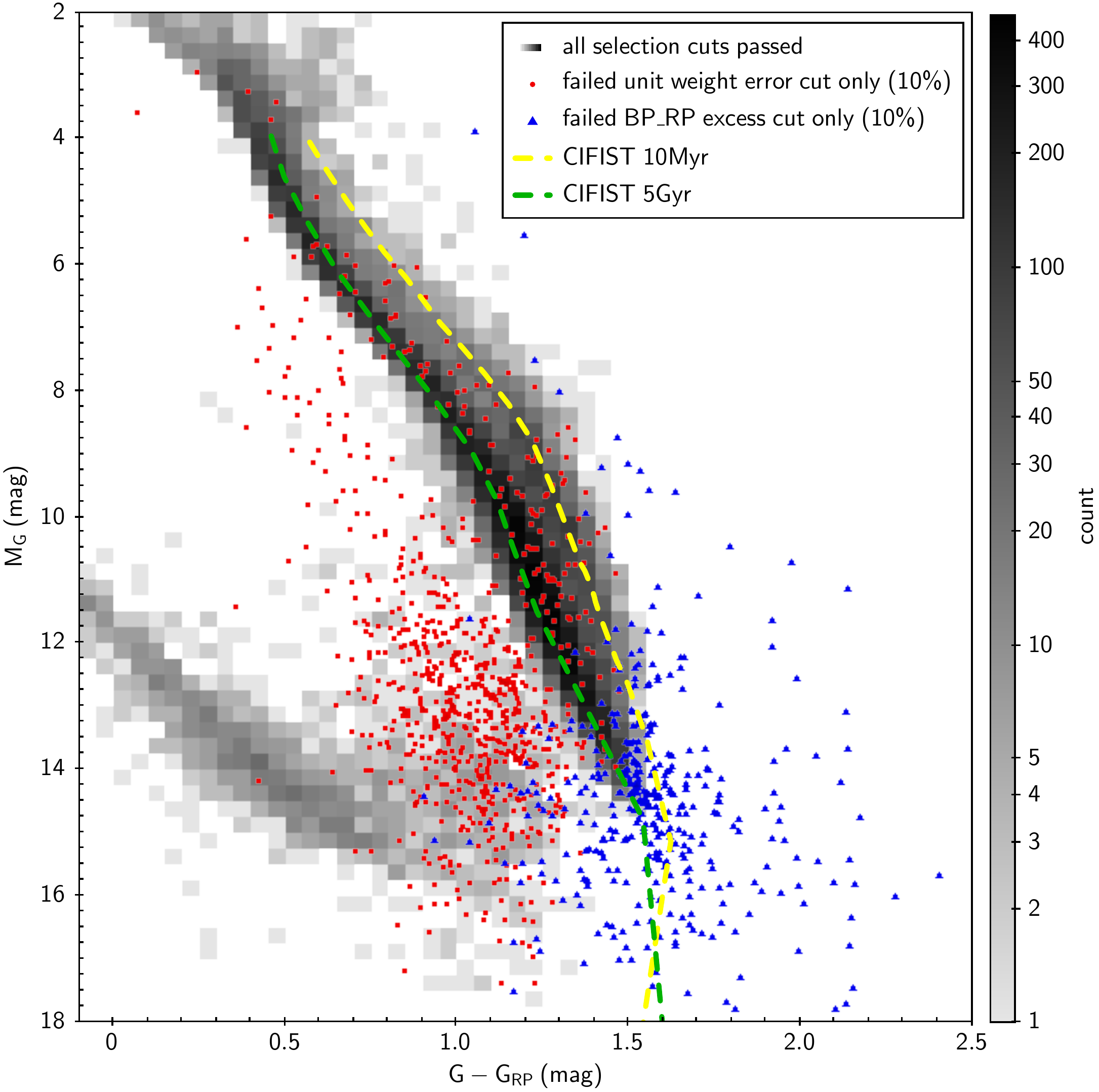}
   \caption{\Gaia\ color-magnitude diagram of all sources in a 20-deg-radius area passing all the quality cuts (2-D histogram with gray scale); except the unit weight error or flux excess ratio cuts (red and blue dots resp., 10\% statistics for readability)}
   \label{cmdcuts}
\end{figure}

For this, we considered all entries in the \Gaia\ catalog over the sky region of interest, with parallaxes corresponding to the expected values of the groups' members,  {i.e. $\varpi>7$\,mas, and any proper motion, in order to have a good statistics}. 
However distant objects entering the nearby sample and hopefully, and rightfully, removed by our selection cuts, should not lead us to underestimate the detection efficiency. To identify them, we used the \Gaia\ color-magnitude diagram, looking for objects on the stellar main sequence to confirm the quality of their measured parallax and photometry. 
The main quality cuts which we applied ({cf. section~\ref{astdet}}) are:
\begin{itemize}
\item ``visibility periods used'' cut: in the magnitude range of interest, the cut {just} removes 1--3\% of the entries. 
Only 2\% of the area has a detection efficiency lower than 70\% and we do not correct for this effect.
\item ``unit weight error'' cut: this cut removes 41\% of the otherwise satisfactory entries, over a wide range of apparent magnitudes. 
Among those stars, 17\% fall on the main sequence (defined between the two 10\,Myr and 1\,Gyr isochrones; see red dots in Fig.\,\ref{cmdcuts}) and may therefore be {\it bona fide} nearby stars with a roughly correct \Gaia\ parallax (but possibly an erroneous uncertainty). As we removed those stars from further analysis, we may have lowered our sensitivity. 
\item ``flux excess ratio'' cut: 27\% of the otherwise satisfactory stars fail this cut, mostly with $G>19$\,mag (see blue dots in Fig.\,\ref{cmdcuts}), causing a severe loss of sensitivity at the faint end. 
Among those stars, 11\% are located on the main sequence.
\end{itemize}
To correct our loss of sensitivity, we considered that the stars removed by either the ``unit weight error'' cut or the ``flux excess ratio'' cut but falling on the main sequence should be added to the nearby sample. 
For each $G$ magnitude bin, we measured the fraction of stars passing all cuts and further analyzed, to the more completed sample derived above. In the mass functions, we increased the weight of each member candidate by that fraction.

In addition to those cuts, we selected stars with \mbox{$\varpi>10\times\sigma_\varpi$}. We can readily determine which fraction of our $\varpi>7$\,mas sample satisfies this cut, with more than 90\% completeness down to $G=19\,$mag, and we corrected for it.  

Finally, one needed to know the completeness of \Gaia\ before any cut is applied. Ideally we would compare with a highly pure, deep optical catalog over the Crux sky region, but there are none. Comparing with the Pan-STARRS1 DR1 catalog \citep{Chamb16,Magni16}, with a high-purity subset of objects with 20 or more detections, in two distinct fields with the same ecliptic latitude as the Crux field: $\epsilon=-45\deg$. We found very different 90\% completeness limits, from $G=20.7$\,mag at a longitude of $129\deg$, to a brighter limit of $G=19.2$\,mag at $90\deg$.  {Instead we turned to the VVV near-infrared survey. Its second data release \citep{Minni17} offers some overlap with the Crux region. We searched for \Gaia\ counterparts to $5.10^5$ VVV stars with null {\tt pperrbits} in the Y and J bands, within~$1\arcsec$. 
\BG{In order to transform the VVV near-infrared photometry to the $G$ magnitude, we derived} the color transformation\footnote{Because the VVV fields cover the Galactic plane with \mbox{$|b_{\rm gal}|<2\deg$}, this relation depends strongly on extinction in the area and stellar sample considered. Here $l_{\rm gal}$=295---318$\deg$.}: 
$$
G_{\rm YJ}=Y_1+2.48(Y_1-J_1)+0.63, 0.4<Y_1-J_1<0.6\,\rm mag,
$$
\BG{We then} determined that the \Gaia\ 95\% completeness limit is fainter than $G_{\rm YJ}=20$\,mag.
}

Combining all those effects, we estimated our 50\% completeness magnitude of $G=19\,$mag, which corresponds to a mass of 0.02\,M$_\odot$ for a 10-Myr CIFIST isochrone. We found no fainter candidates and did not attempt to constrain the mass function for lower masses. 

\begin{figure}[htb!]
  \centering
  \plotone{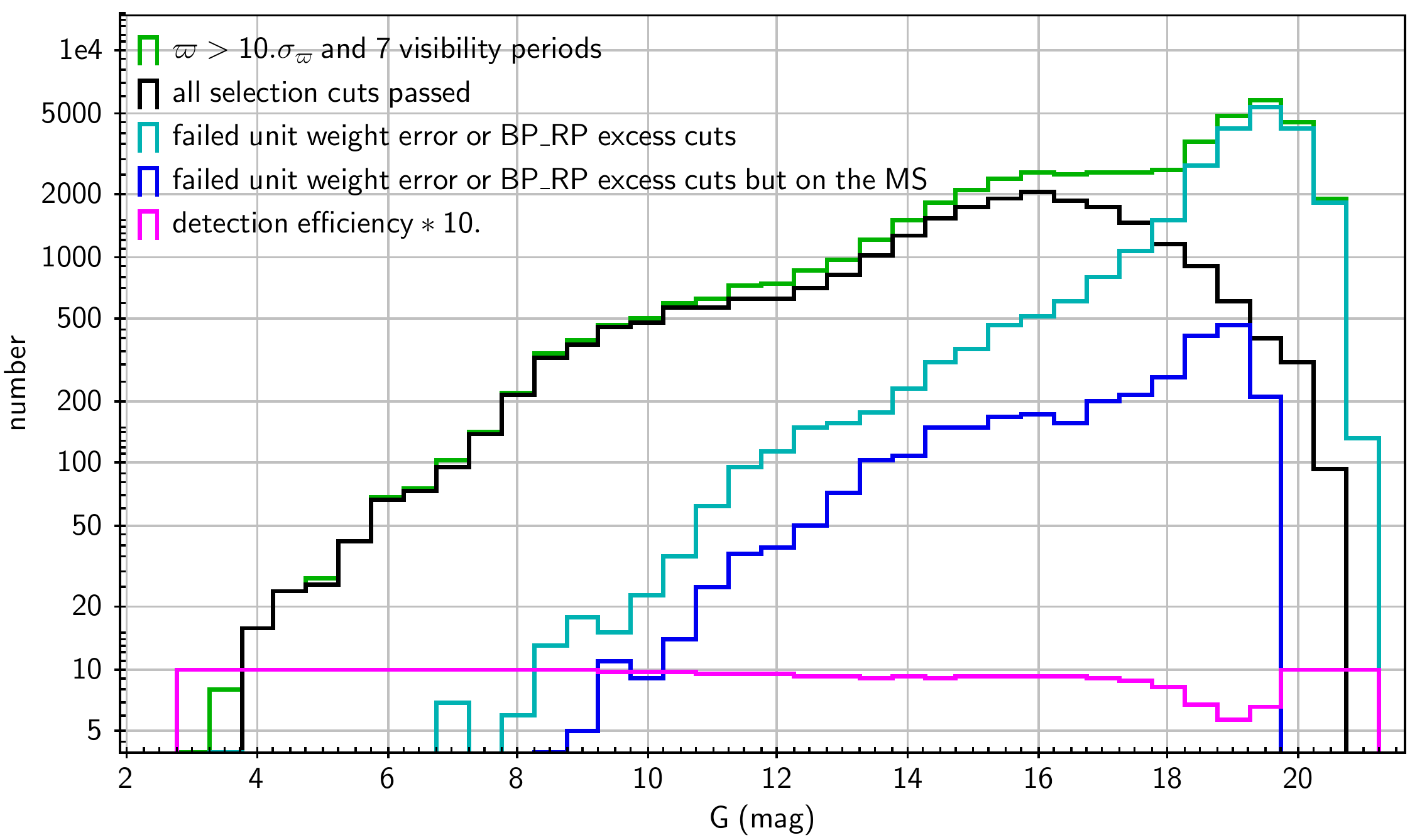}
  \caption{Total number of \Gaia\ entries in a 20-deg-radius area with $\varpi>7$\,mas (red), and $\varpi>10\sigma_\varpi$ and 7 visibility periods or more (green). The pale blue histogram show all the sources passing all the quality cuts (2-D histogram with color scale) except the ``unit weight error'' (top-left panel) or ``flux excess ratio'' cut (top-right); while the blue histogram shows among those sources, those that are located on the main sequence (see Fig.\ref{cmdcuts}). }
  \label{histocuts}
\end{figure}
   
 {The kinematic selection may also lead to the loss of some CMG binaries, when their orbital motion is comparable or larger to the selection box size in the tangential velocity plane (see Fig.\ref{V_par_sen_young}). For face-on, circular orbits, this will affect systems with a total mass $M_{\rm total}/{\rm M}_\odot>10^{-2}a$ where $a$ is the semi-major axis in A.U.
Following \citet{{Duche13}}, this may affect a significant fraction of stars later than F, which have typical semi-major axis of tens of A.U. Early-type stars, with semi-major axis less than 1\,A.U., may not be well fitted by the single-star astrometric model, and have been removed from the \Gaia\ catalog.}

\subsection{Contamination}   

In the paragraph above we discussed the completeness of our sample using the observations
of \Gaia\ DR2. On the other hand, we also had to estimate a possible contamination to CMG by
"field" stars. Since we selected the 1844 stars of CMG within a narrow range of tangential
velocities (see Fig.~\ref{V_par_sen_young}), the contamination should be rather low. However, we
cannot exclude the possibility that there are, just by chance, a few young stars with the appropriate  $v_\parallel$ and $v_\bot$ velocities but with an inadequate third component, the radial velocity.
Provided that such stars have a roughly uniform distribution in
Fig.~\ref{V_par_sen_young}, we derived the number of contaminants by comparing the
average number density of stars in Fig.~\ref{V_par_sen_young} in- and outside the area of CMG. The 1844
stars of CMG cover an area of 11 $(\rm km~s^{-1})^2$. For the comparison we selected an               
area of 30 $(\rm km\,s^{-1})^2$ centered at $(26.5,0)\,\rm km\,s^{-1}$, where we find 47 stars that we rated as a typical
background in Fig.~\ref{V_par_sen_young}. This gives a possible contamination of 17 stars to our 1844 or about 1\%. So, this has insignificant impact on the mass function of the CMG.   

\subsection{Mass functions} \label{MFs}
{
As stated in Sec.~\ref{massdet} we determined masses of individual members by using the isochrones by \citet{2017ApJ...835...77M} for masses higher than 0.7\,M$_\odot$. For lower masses we took the isochrones by \citet{2015A&A...577A..42B} as they model quite well the shape of the observed stellar and sub-stellar sequences for our objects in CMG.
Making corrections for incompleteness (Sec.~\ref{deteff}), we determined the mass functions for all four groups, and for the inter-group members. They are shown in Fig.\ref{MF}, top panel. \BG{We caution that unresolved binaries are counted once, and did not attempt to estimate the system total mass out of the mass fitted on the combined photometry. We studied the case of resolved binaries in more details in Section\,\ref{ResBins}.}

We fitted a log-normal function \citep{Scalo86} to our data, and present the mass functions of \citet{Salpe55}, \citet{Chabr03a}, and \citet{Kroup02} for comparison (see Fig.\ref{MF}, bottom). We found that all the mass functions of our sub-groups are very similar and well reproduced by a single log-normal function of mean mass $m_c=0.22$\,M$_\odot$ with a dispersion of 8\%, and a very consistent standard deviation {$\sigma=0.64$, hence slightly wider than the canonical \citet{Chabr03a} mass function (for systems) with $\sigma=0.55$}.
In particular, we found no difference between the groups' mass functions and that of the inter-group members. Hence, the latter are not only low-mass members ejected by dynamical interactions. There is an excess at about 1.4 M$_\odot$ which may have been caused by an imperfect separation between young and old stars.

\begin{figure}[htb!]
  \centering
  \plotone{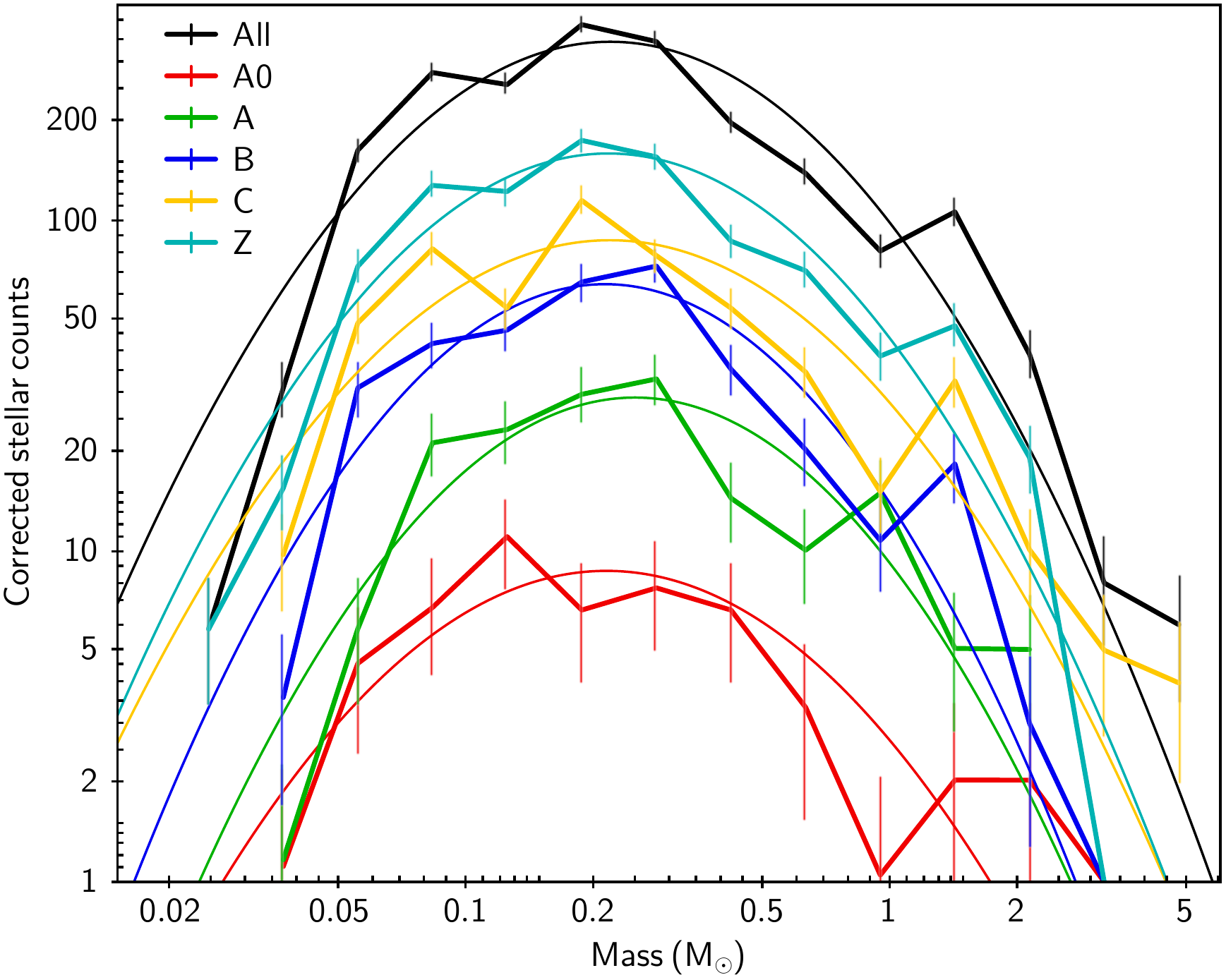}
  \plotone{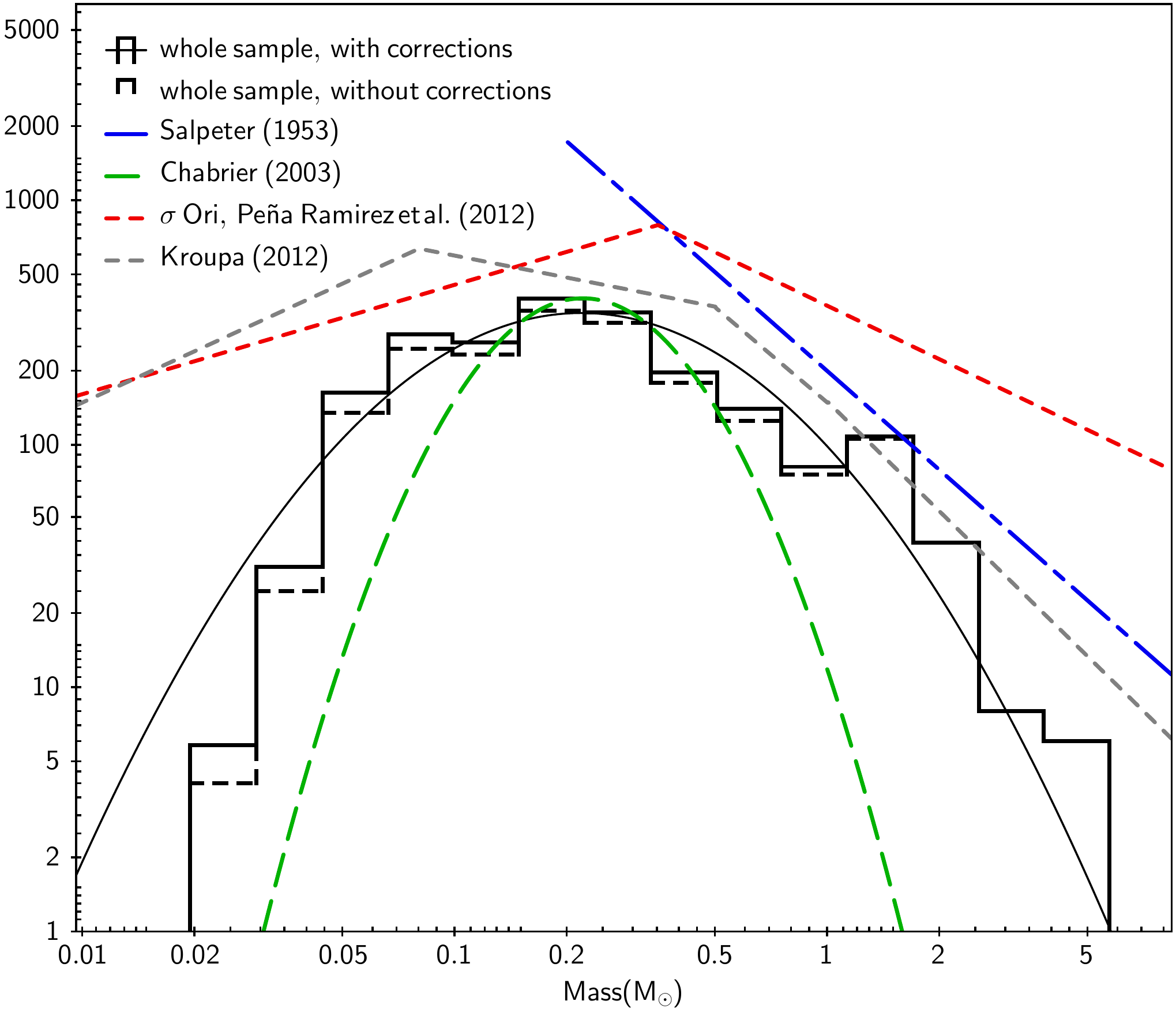}
  \caption{Top: Mass functions of all four groups (in color), the intergroup members (Z) and of the whole cluster (black), together with a log-normal fit. Bottom: Mass function of the whole cluster before (dashed histogram: stellar counts) and after (solid histogram) correction for incompleteness. We include various canonical mass functions, which we did not normalize for readability, as well as the power-law fit of the $\sigma$~Ori cluster \citep{PenaR12}.}
  \label{MF}
\end{figure} 

The added-up masses of the members detected in each group are listed in Table\,\ref{TabGroups}. The total mass of the stars and brown dwarfs in CMG is 673\,M$_\odot$.

We can compare our mass function to the mass functions reported for other young clusters or associations.
\citet{PenaR12} also used the isochrones from the Lyon group, NextGen98 \citep{1998A&A...337..403B}, AMES-Dusty \citep{2000ApJ...542..464C}, and AMES-Cond \citep{2003A&A...402..701B}, to determine the low-mass part of the mass function of the $\sigma$\,Orionis cluster with an age of 2 to 4 Myr. They slightly favor power laws over a log-normal form, with $\alpha=0.6$ for masses less than 0.35\,M$_\odot$,  {where $\alpha$ is the index of the power law: $\Delta N/\Delta M=M^{-\alpha}$}. Their log-normal fits in this mass range have $\sigma$ between 0.57 and 0.64, so are comparable or slightly smaller than our value. They note however that this mass function considerably underestimates the number of planetary mass objects ($m<0.012\,\rm M_\odot$), which is outside the mass range we are covering in this paper. While $\sigma$\,Orionis is a very compact cluster with a characteristic radius of only 1~pc \citep{Bejar11}, CMG is a more extended group of young objects. So, the star formation mode of both  may differ. 

\citet{2013MNRAS.431.3222L} studied the Upper Scorpius part of the Scorpius-Centaurus association, and found that for masses higher than 0.03\,M$_\odot$ is well reproduced by the log-normal form of the Chabrier mass function, independent of the age one determines for Upper Scorpius. This is a steeper decline from the $m_c=0.22$\,M$_\odot$ towards 0.03\,M$_\odot$ than we found for the neighboring Lower Centaurus Crux (LCC) part. As \citet{PenaR12} for
$\sigma$\,Orionis, \citet{2013MNRAS.431.3222L} finds an excess of objects with masses lower than 0.03\,M$_\odot$ depending on the adopted age. Again this is beyond the scope of this paper.

Very recently, \citet{2017MNRAS.471.3699M} studied the deeply embedded young ($\approx$ 1 Myr), massive and very dense star cluster RCW 38, and found $\alpha = 0.71 \pm 0.11 $ for masses between 0.02 and 0.5 M$_\odot$, a decline towards low masses even shallower than in $\sigma$\,Orionis.

Summarizing, we found that the decline of our mass function from the mean mass of 0.22\,M$_\odot$ down to 0.03\,M$_\odot$ is steeper than that found for the $\sigma$\,Orionis cluster and RCW 38, but shallower than in the case of Upper Scorpius found by \citet{2013MNRAS.431.3222L}.}

\subsection{Age distributions} \label{Ages}

{
As mentioned above, we simultaneously determined the model masses and ages of individual members. For the mass and age determination for low-mass stars and sub-stellar objects we relied on the CIFIST isochrones. Firstly, because they model the observed loci of the object quite satisfactorily in all \Gaia\ passbands, whereas in the case of the Parsec 3.0 isochrones, the magnitudes in the $G_{RP}$ passband come out way too faint. This leads to the situation that ages from Parsec 3.0 are lower than the CIFIST ones, when $G_{RP}$ is included and higher, when it is not. The CIFIST isochrones for ages between 5\,Myr and 40\,Myr are mostly parallel to each other in the \Gaia\ color-magnitude diagram.

In Fig.\ref{SFR}\,(top) we show the age distribution for each sub-group. The bulk of the ages is between 5\,Myr and 20\,Myr with few nominally older objects (note the logarithmic scale here). There are slight differences in the average ages of the subgroups of CMG (shown in Table\,\ref{TabGroups}), increasing from as low as 7\,Myr in A0 to 10\,Myr for group C and the inter-groups members. 

   \begin{figure}[htb!]
   \centering
   \plotone{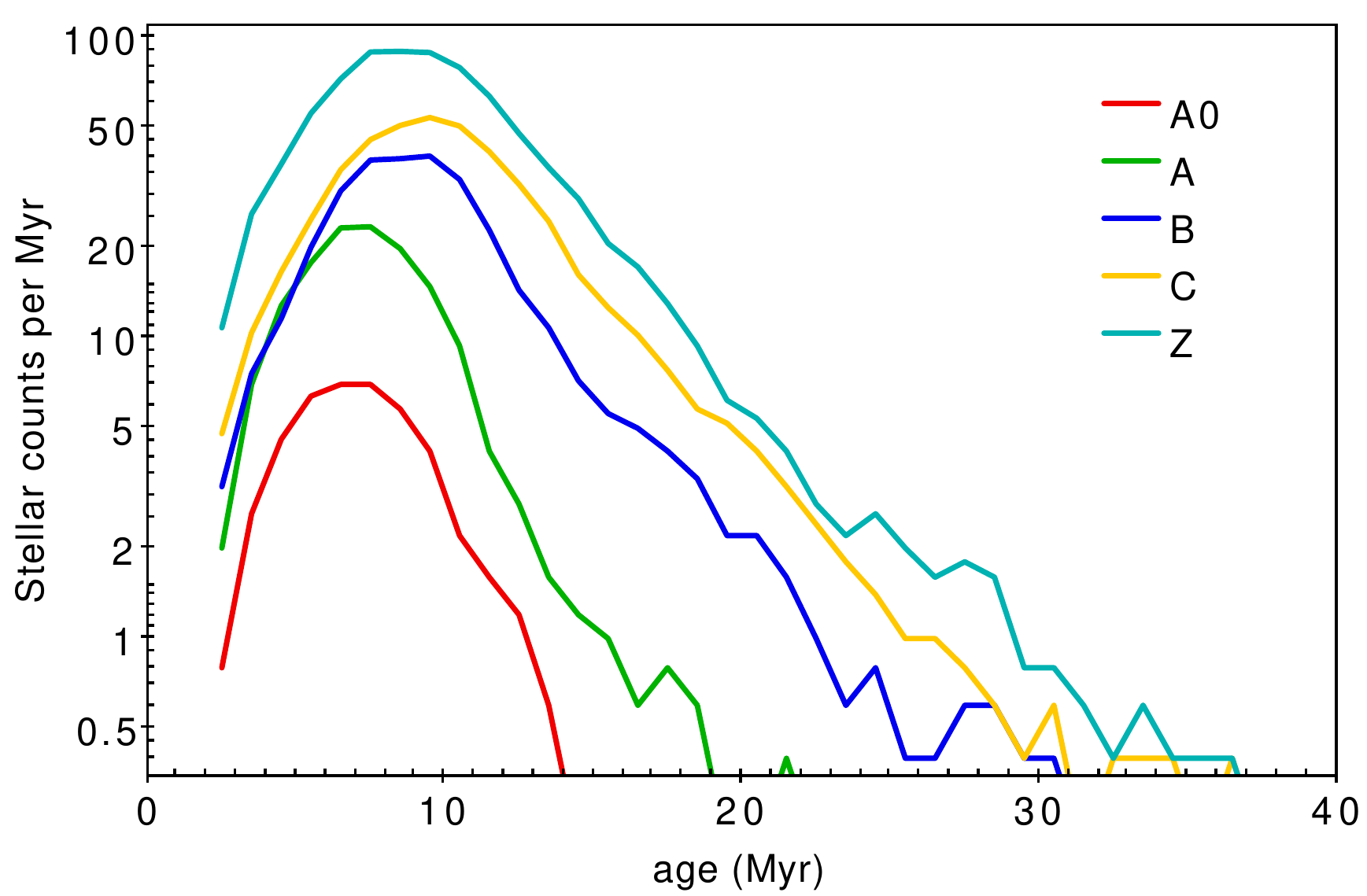}
   \plotone{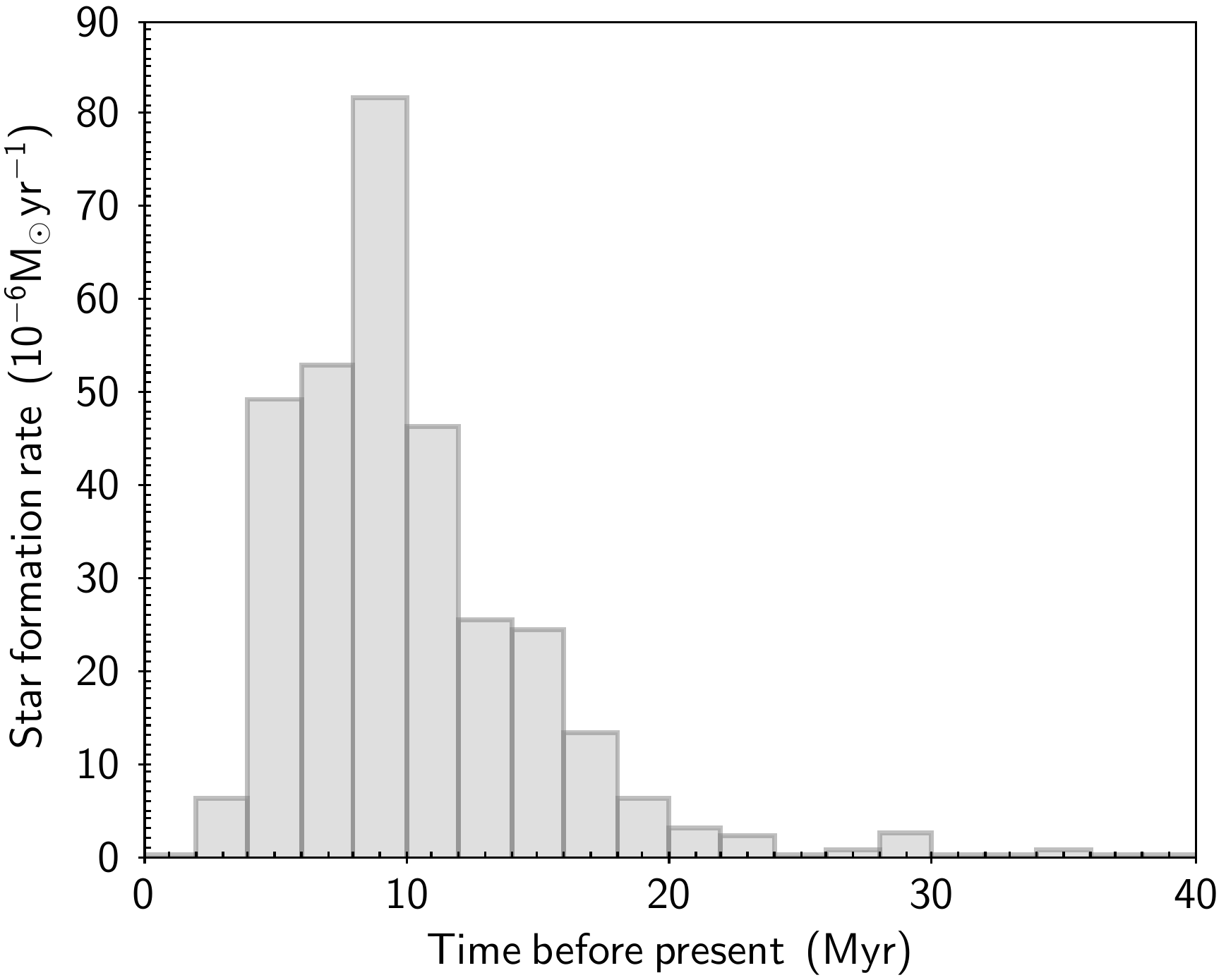}
      \caption{Top: CIFIST age distribution for each subgroup and the inter-group sample (Z). We stress that the stellar counts, measured for bins of width 5\,Myr, are not independent. Bottom: The star formation rate in the CMG as a whole as a function of time. }
         \label{SFR}
   \end{figure}
   
Fig.\ref{SFR}\,(bottom) shows the star formation rate in the CMG. We do not distinguish here between the sub-groups. Star formation essentially started 20 Myr ago and reached its maximum
between 8 and 10 Myr before present with a rate of \mbox{$8\times10^{-5}$ $\rm M_\odot\,yr^{-1}$}. 

\BG{Those ages are similar to those reported by \citet{2012AJ....144....8S}, derived from Li $\lambda 6708$ absorption strength analysis. 
These authors determined a mean age of 10\,Myr, without reporting individual ages.
We note that we recovered 84\% of their sample over the area we studied (see Table\,\ref{others}). 
While their results are model dependent, just like ours are, the two methods are independent and, if any, would have different biases. By comparing the lithium absorption strength,  \citet{2012AJ....144....8S} derived a younger age for their LCC targets than for the $\beta$\,Pictoris moving group members, which they estimated at 12\,Myr.

We can compare on an individual basis our ages with those of \citet{2016MNRAS.461..794P} for the sample in common of 95 stars.
They agree well to within 15\% on average for the low-mass stars (our mass $m<0.8\,\rm M_\odot$), albeit with a large dispersion. However the ages of \citet{2016MNRAS.461..794P} for higher mass stars ($m>1\,\rm M_\odot$) are 30\% larger than ours. In fact, \citet{2016MNRAS.461..794P} detected a correlation between effective temperature and  mean ages, while we did not.

On the other hand, for} the young population in LCC, \citet{Mamaj02} found mean ages of the pre-main sequence stars
from Hipparcos and TYC/ACT to range between 17 and 23 Myr. When using the 	
\citet{1994A&AS..106..275B} tracks, they determined main-sequence turnoff ages for Hipparcos B-type members to be 16 $\pm$ 1 Myr. It is worth mentioning that much earlier \citet{1989A&A...216...44D} and \citet{1985ASSL..120...95D} obtained turn-off ages between 10 and 12\,Myr, so much closer to the ages determined here. The difference between these earlier attempts to get an age for LCC and ours lies in the availability of \Gaia\ DR2. 
\BG{Indeed, we detected a clear correlation ($R\geq 0.6$) between the relative offset of the kinematic parallaxes used in \citet{Preib08} and \citet{2016MNRAS.461..794P} to the \Gaia\ DR2 parallaxes, and the relative offset of the ages derived in those articles, to ours. An overestimated parallax leads to an underestimated absolute luminosity and to an overestimated age.
In addition, the majority of our} members has masses way below those that could be used in the papers above, and the photometric quality is much superior. }

\section{Remarkable members} \label{members}

\subsection{Circumstellar disks} \label{disks}

With ages below 20\,Myr, a fraction of our CMG members is likely to harbor a circumstellar disk at various stages of evolution.

We have used the Virtual observatory SED analyzer \citep[VOSA][]{Bayo08} to analyze the spectral energy distribution of our members. 
 {In a nutshell, VOSA adjusts theoretical isochrones available in the Virtual observatory (VO) to the photometric information provided by the user and/or collected on the VO, by $\chi^2$ minimization. It detects deviations from the photospheric emission, such as the infrared excess of interest here, and ignores the corresponding data points when the fit is repeated.}
We used all available VO photometry and fitted the data with the CIFIST models  {with Solar metallicity} \citep{2015A&A...577A..42B}, refining once the infrared excess detection. 
\BG{For A- and B-type stars, which the CIFIST grid does not cover, we used the ATLAS9 atmosphere models \citep{Caste97}, again with solar metallicity and \mbox{$E(\rm B-V)=0$}.}
{The best model parameters are listed in Table\,\ref{memtable}.} We then selected for further inspection those stars with an excess of at least 0.25\,mag and $5\,\sigma$, in at least one filter. We removed by eye a fraction of the candidates with clearly poor fits. Finally we combined the official quality flags provided by the AllWISE catalog \citep{2010AJ....140.1868W} and the results of a careful visual inspection of the relevant WISE images.

\subsubsection{Identification of spurious WISE detections}

 Most of our excess candidates are located within 10{\degr} of the galactic
 plane. The higher source density as well as the bright extended background
 emission results in enhanced contamination and spurious detections in this
 region \citep{koenig2014}. We took special care to identify and discard
 candidate sources affected by such issues. The AllWISE catalog already  provides
 useful information on the data quality. We removed all targets where  detection
 in any of the four bands was qualified to be spurious either because  of a
 diffraction spike ("D" flag),  scattered light halo ("H"), optical ghost  image
 ("O") or latent image ("P") from a nearby bright source based on the ``confusion
 and contamination flag" parameter of the AllWISE catalog.  Sources found to be
 extended in the W3 or W4 bands (i.e. the reduced $\chi^2$  of their profile-fit
 photometry, {\tt w3rchi2} or {\tt w4rchi2} is  higher than 3) were also
 discarded.  Furthermore, to ensure good quality data,  only photometry with
 signal-to-noise  ratio higher than 5 was kept in our analysis. According to
 \citet{koenig2014}, especially in regions with high sky  background, a
 significant fraction of spurious detections may have remained  unrecognized in
 the AllWISE catalog. To identify additional potentially problematic  cases we
 inspected the WISE images of our excess candidates visually. This process revealed
 several additional fake sources, and also objects whose photometry is seriously
 contaminated  by nearby bright sources or background nebulosity. {The applied quality criteria reduced the size of our sample to 170~systems. For 136 of them both the W3 and W4 bands data found to have good quality, for 34~objects the W4 band photometry was not used in the following analysis.}

 \subsubsection{Classification of disk candidates} \label{sec:diskclassification}

\begin{figure}[htb!]
  \centering
  \plotone{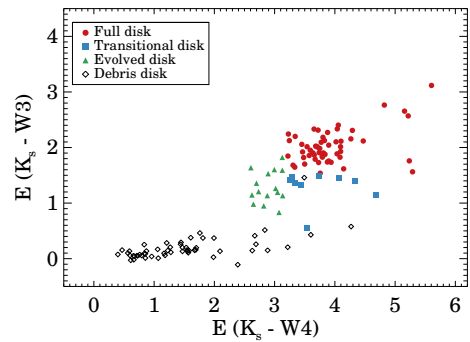}
  \caption{Color excesses for disk candidates having good quality 
   photometry both in W3 and W4 bands. Classification of the disks 
   is based on criteria proposed by \citet{2012ApJ...758...31L} and
   \citet{2016MNRAS.461..794P} (for 
   details, see Sect.~\ref{sec:diskclassification}).}
  \label{fig:diskclassification}
\end{figure}

 The observed infrared excesses, outlined by the WISE photometry, indicate the presence of circumstellar disks around our 170\,selected stars. To assess the evolutionary stage of the revealed disks we
 applied the classification scheme proposed by \citet{2012ApJ...758...31L} and
 \citet{2016MNRAS.461..794P}. This scheme allows to distinguish four different classes:
 full disks,  transitional disks, evolved disks, and debris disks \citep[for
 details of the different classes, see][]{2012ApJ...747..103E} based on the $E(K_s-W3)$
 and $E(K_s-W4)$ color excesses of the objects.  To compute these color excesses
 the predicted photospheric $Ks-W3$ and $Ks-W4$ colors  were taken from the VOSA
 models described above. Where no good quality W4 photometry was available the
 full, transitional, and evolved disk categories could not be distinguished
 unambiguously based on the applied criteria. For these cases we could discern
 three different categories based on the  $E(K_s-W3)$ excesses: full or evolved
 disks, transitional or evolved disks, and debris disks.  In three special cases
 the resulting classification was reconsidered based on literature data.
 Using the above scheme \object{HD\,110058}, 
\object{HD\,121191}, 
\object{HD\,121617} 
were  classified as primordial disks. 
However, though these objects exhibit strong mid-infrared excess and harbor detectable amount of CO gas \citep{lieman-sifry2016,moor2017}, they {were} generally categorized as debris disks based on their spectral  energy distributions and dust masses \citep{melis2013,lieman-sifry2016,moor2017}. 
Considering these more detailed
 studies the three targets finally have been classified as debris disks.
 Color excesses as well as the obtained classifications are displayed in
 Fig.~\ref{fig:diskclassification}.
{As a result of our analysis we identified 115 primordial (full disks, transitional disks, or evolved disks) and 55~debris disks in our sample. }
 {The objects having disks are listed in the \BG{table available online and described in Table\,\ref{memtable}, along with their} main Simbad name and the discovery reference.}

\BG{We note that 15 debris disks of the Sco-Cen association listed by \citet{chen2011,chen2012}, are not recovered by our search. Nine are not in our CMG sample, generally because of poor \Gaia\ data, an isochrone age larger than 40\,Myr, or because their parallax is smaller than 7\,mas. }

\subsection{Brown Dwarfs}

 {Given the small distance of the LCC and its young age, \Gaia\ is able to detect even low-mass brown dwarfs in this region.}
Among the 1844 objects in the CMG sample there are  {214} with masses smaller than  {0.073}\,M$_\odot$, i.e. brown dwarfs.
Only for one of them, \object{[FLG2003] eps Cha 16}, we found an entry in SIMBAD. This young stellar object (YSO in SIMBAD)
has spectral type M5.75 (we determined $M_G = 11.51$, mass = 0.05 M$_\odot$, age = 7\,Myr), is located at the farthest edge of the CMG sample, and is a member of our A0 Subgroup.  
War
Among those 214 brown dwarfs, 34 harbor a circumstellar disk. While we are not complete with respect to transitional and debris disks that are too faint to be detected with our WISE selection, we note than ($14\pm 3$)\% of the brown dwarfs bear full disks or full/evolved disks. This is eight times higher (a $5\sigma$ effect) than for stars  with a fraction of ($1.6\pm 0.5$)\%, possibly meaning that brown dwarfs keep their disk optically thick at infrared wavelengths longer than stars.

\subsection{Resolved binaries} \label{ResBins}

The \Gaia\ DR2 catalog has a nominal spatial resolution of 0.4\,arcsec, i.e. 46\,A.U. at the mean member distance. 
According to \citet{Duche13}, about half the stellar binaries made of components with masses between 0.7 and $5\,\rm M_\odot$ have semi-major axis larger than that separation.
These binaries, through their additional orbital motion, perturb the study of the internal dynamics of the cluster; and may lead to incompleteness (see Section\,\ref{deteff}). 
\BG{In addition, both components of resolved binaries would be counted separately in luminosity and mass functions, as those that we analyzed in Section\,\ref{MFs}, while unresolved binaries are counted only once, so it is important to identify resolved binaries in order to derive a coherent luminosity and mass functions}.
Finally, the orbits of high-mass systems with semi-major axis of a few tens of A.U. will ultimately be determined by \Gaia, and therefore their dynamical masses of the components will be measured.

We searched our member catalog for pairs sharing the same proper motions and parallaxes.
To estimate the contamination by random pairing, we used the standard procedure of creating a mock catalog of stars derived from the observed sample, shifting one coordinate by some angle, so that any pairing between the observed and mock catalogs can only be a spurious match \citep[see][for an illustration of the method]{Lepin07}. To estimate the quality of the match, we calculated the sum $\Sigma$ of the differences between the proper motions and parallaxes, expressed in $\sigma$.

We include in the error budget the orbital motion, which will introduce a difference in the proper motions, calculated for a face-on, circular orbit.

We selected observed pairs up to 10\,arcmin and $\Sigma<9$. For each pair, we determined the number of spurious pairs with parameters (sky separation, $\Sigma$) smaller than the observed pair, i.e. better matches. To improve the statistics, we shifted the observed sample either RA or DEC by 0.5, 1, and 1.5\degr.
We identified 8\,pairs, with separation up to 3\arcmin, for which the number of better spurious matches is smaller than 2.2. 
They are listed in Table\,\ref{pairs}.
The next best candidate, with a separation of 400\arcsec and $\Sigma=7.2$, have 20 better spurious matches. 

Although the statistics is limited, we find that the mass ratios increase from 0.1--0.2 for F, G stars to almost unity at the end of the main sequence, as has been found in previous studies \citep[for a review, see][]{Duche13}. The lowest-mass pair we identified has a total mass of 0.4\,M$_\odot$, and the closest separation is 3\,arcsec.

\begin{table*}[htb!]
\centering
\caption{Binary members.}
\begin{tabular}{l c c l c c c c}
\hline\hline
  \multicolumn{1}{c}{Gaia ID 1} &
  \multicolumn{1}{c}{Group 1} &
  \multicolumn{1}{c}{Mass 1} &
  \multicolumn{1}{c}{Gaia ID 2} &
  \multicolumn{1}{c}{Group 2} &
  \multicolumn{1}{c}{Mass 2} &
  \multicolumn{1}{c}{Separation}  &
  \multicolumn{1}{c}{Contaminants}  \\ 
  \multicolumn{1}{c}{} &
  \multicolumn{1}{c}{} &
  \multicolumn{1}{c}{($\rm M_\odot$)} &
  \multicolumn{1}{c}{} &
  \multicolumn{1}{c}{} &
  \multicolumn{1}{c}{($\rm M_\odot$)} &
  \multicolumn{1}{c}{(arcsec)}  &
  \multicolumn{1}{c}{}  \\ 
\hline
6075310478057303936 & C &  0.75 &    6075310478050174592 & C &  0.24 &  3.232 &  0 \\
6065138792911302784 & Z &  1.62 &    6065138792911300096 & Z &  0.13 &  19.413 &  0 \\
5861825550891696000 & A &  0.20   &   5861825349107118976 & Z & 0.19 &   31.073 &  0 \\
6065047911406764288 & Z & 1.37 &   6065047636528853888 & Z & 0.26 &   51.768 &  0 \\
5894194318558985984 & Z &  1.17 &   5894194348576228992 & Z & 0.66 &   59.699 &  0.2 \\
6087959014999185152 & Z &  0.92 &    6087958675701081600 & Z & 0.10 &  113.306 & 0.7 \\
6090729754245639552 & Z &  0.39  &   6090730132201936256 & Z & 0.18 &   149.843 & 1.2 \\
5854812388999707008 & Z & 1.60 &   5854813312408616320 & A0 & 0.30 &   166.388 & 3.4 \\
\hline\end{tabular}
\label{pairs}
\end{table*}

\section{The motion of CMG}\label{motion}

\EuS{
Table~\ref{TabGroups} gives the full 6-D average phase space coordinates for the 4 sub-groups of GCYS in the Galactic Cartesian coordinate system. These coordinates are determined exclusively from the observations in Gaia DR2, i.e. only for those members that have measured radial velocities in DR2. In Table~\ref{TabGroups} we find
correlations between the spatial coordinates and the respective velocities of the sub-groups. This suggests an expansion of the whole CMG. The good accuracy of the mean space velocity components (median 0.2 km\,s$^{-1}$) allows to trace the groups back in time to find the minimum distance of the groups from each other. Given the ages of the stars in CMG, we trace them back  for not more than about 20\,Myr. In the Galactic Cartesian coordinate system we used a simple 3-D linear approach $\vec{X}(t) = \vec{X}(0)+t\times \vec{V}$ with the components of the space vector $\vec{X}$ in parsec and the components of the velocity vector $\vec{V}$ in pc\,Myr$^{-1}$ (1 km\,s$^{-1}$ = 1.0227\,pc\,Myr$^{-1}$).
We traced back the centers of the subgroups from Table~\ref{TabGroups} in steps of 0.1\,Myr using the data given in this table.
To characterize the size of the CMG, we used the quantity $ R = \sqrt {\sum \limits_{i=0}^{N} \sum \limits_{j=i+1}^{N}r_{i,j}^2}$ where r$_{i,j}$ is the 3-D distance between sub-group $i$ and subgroup $j$. We found that $R$ had a minimum at $-10$\,Myr. That means that the centers of the 4 sub-groups had its closest distance from each other $10$\,Myr before present and are drifting away from each other since. This finding is compatible with the ages estimated for the individual stars in CMG.

In this connection the question arose if the sub-groups we see now, were also separated from the beginning, or if the CMG is expanding as a whole. To test this hypothesis it is necessary to trace back the individual stars of the CMG sample, which means we need accurate radial velocity measurements for a representative subset of stars in CMG. In \Gaia\ DR2 \citep{2018arXiv180409372K} we find observed radial velocities for 247 stars of the CMG. The formal precision of the radial velocity measurements varies between 0.2 and more than $10\,\rm km~s^{-1}$ with a median of $2\,\rm km~s^{-1}$, which is at least one order of magnitude inferior to the tangential velocities. To reduce the influence of the poorly measured radial velocities we considered only stars (123 in number) having $\sigma_{V_{r,obs}}$ better than $2\,\rm km~s^{-1}$ (median of $1\,\rm km~s^{-1}$) for the back-tracing. The lower accuracy of the radial velocities influences primarily the $Y$ coordinate, because it is closest to the line-of-sight.  To a lesser extend this holds for the $X$ coordinate, while the $Z$ coordinate is little affected. 

As in the case of the sub-groups we also find correlations between the spacial coordinates and the respective velocity components of the 123 stars. In Fig.~\ref{regression} we show the velocities $U,V,W$ versus the coordinates $X,Y,Z$.
 A linear fit between coordinates and velocities delivers slopes of $0.087 \pm 0.008$ in $X$, $0.064 \pm 0.016$ in $Y$, and $0.067 \pm 0.004$ km\,s$^{-1}$\,pc$^{-1}$ in $Z$. All the slopes are significantly positive which indicates expansion in all three spacial directions. As expected the expansion rate is worst determined (4\,$\sigma$) in the $Y$-direction reflecting the lesser accuracy of the radial velocities.

\begin{figure}[htb!]
  \centering
  \plotone{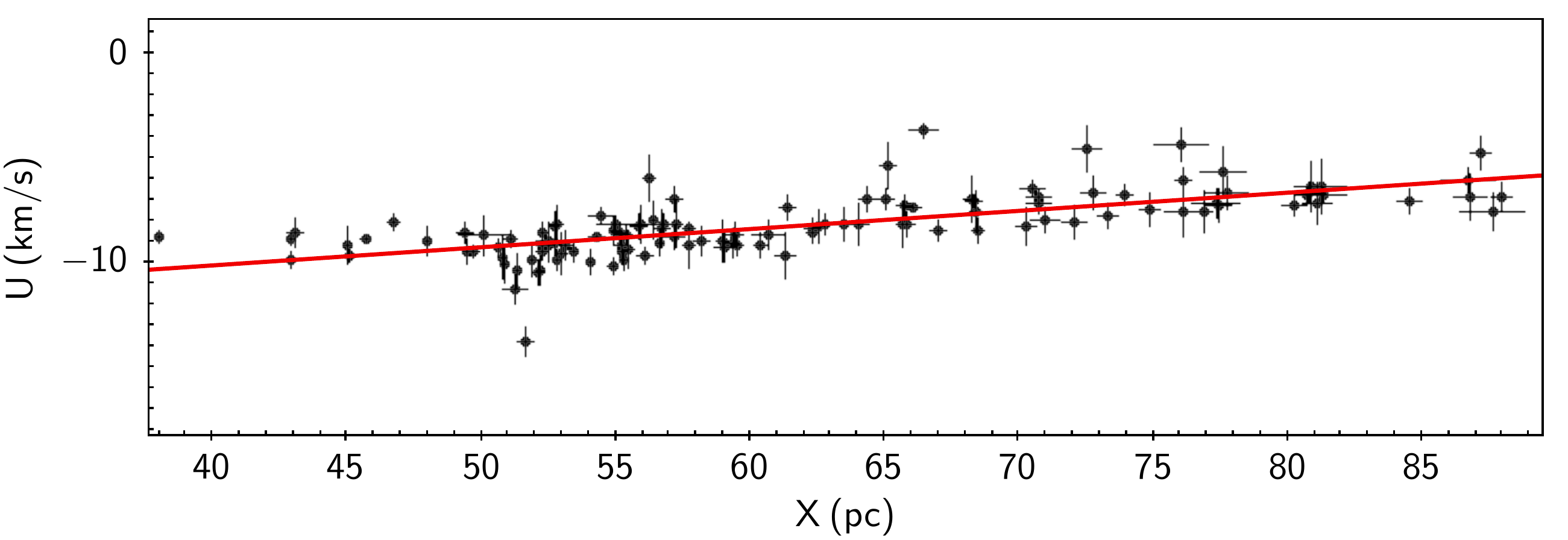}
  \plotone{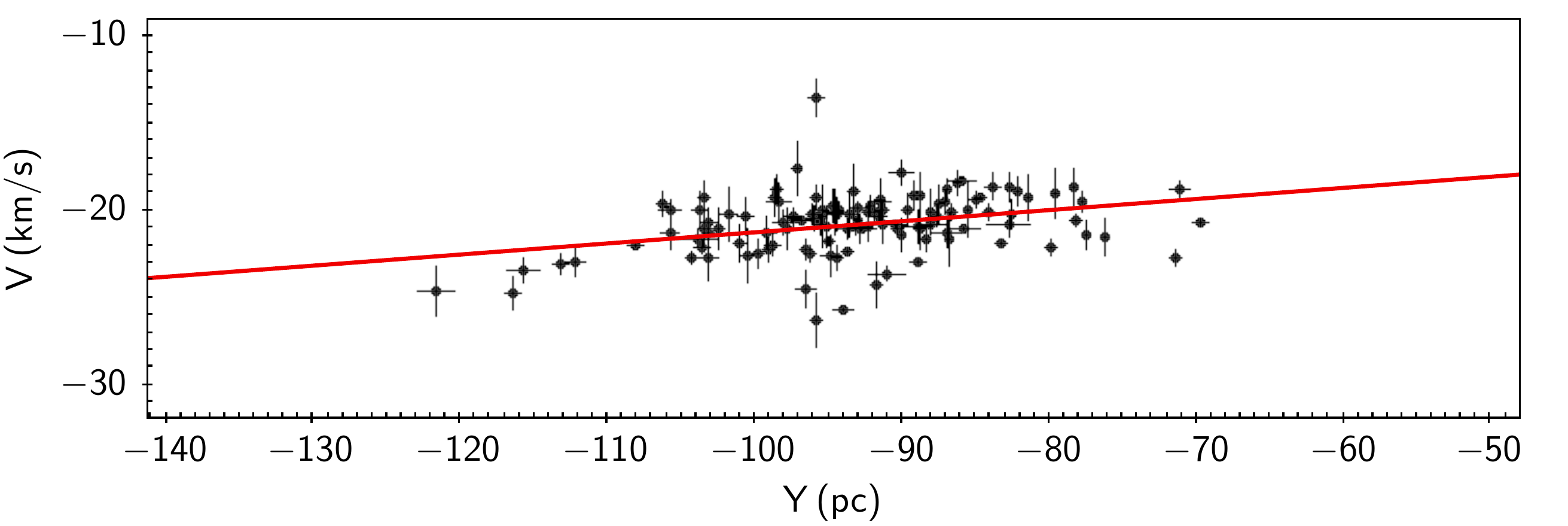}
  \plotone{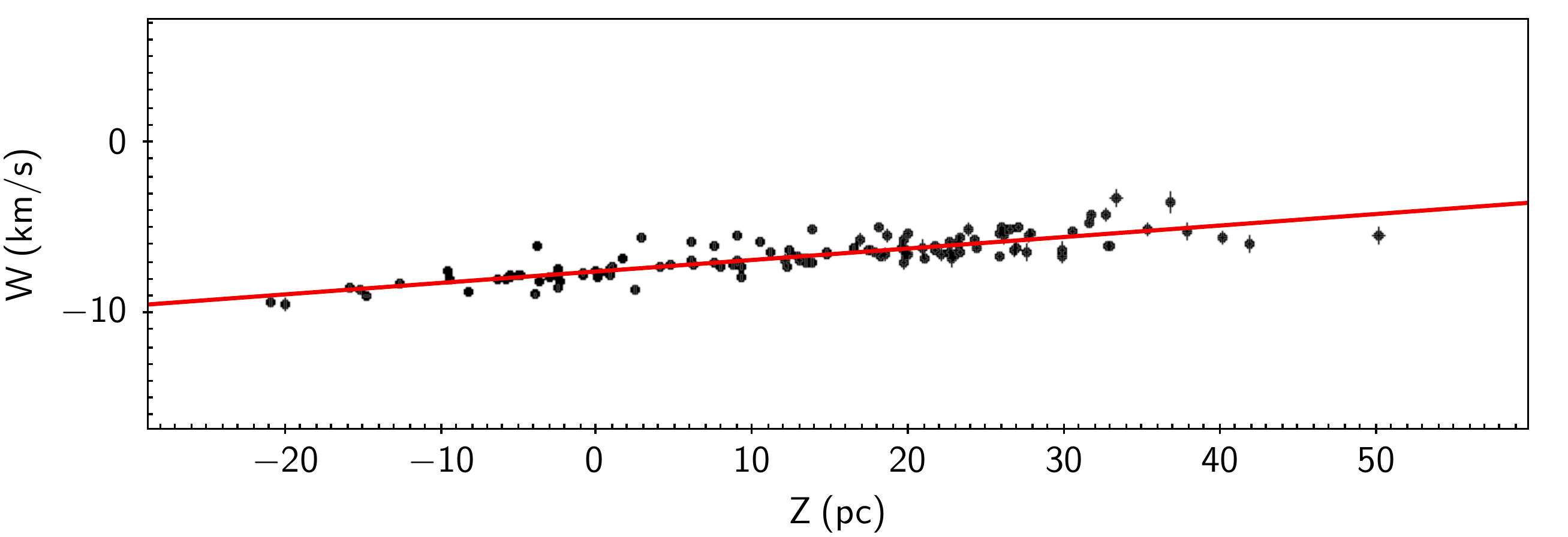}
  \caption{Galactic Cartesian velocities $U,V,W$ versus coordinates $X,Y,Z$ (from top to bottom) for 123 stars in the CMG with radial velocities having $\sigma_{V_{r,obs}}$ better than $2\,\rm km~s^{-1}$. The regression fits with slopes of \mbox{$0.087 \pm 0.008$}, \mbox{$0.064 \pm 0.016$} and \mbox{$0.067 \pm 0.004$\,km\,s$^{-1}$\,pc$^{-1}$} are shown in red.}
  \label{regression}
\end{figure}

Knowing that GCYS is expanding, we performed a full 3-D back-tracing for the 123 stars in the Galactic Cartesian coordinate system to determine the time of minimum extension. We evaluated the quantity $R$ which we re-write  as:\\
$ R^2 = \sum \limits_{i=0}^{N} \sum \limits_{j=i+1}^{N}X_{i,j}^2+\sum \limits_{i=0}^{N} \sum \limits_{j=i+1}^{N}Y_{i,j}^2+\sum \limits_{i=0}^{N} \sum \limits_{j=i+1}^{N}Z_{i,j}^2$\\
with $X_{i,j}^2 = (X_i - X_j)^2$ and so on. This allowed to study the influences of the motions along the $X,Y,Z$-axes onto the $R$. At t=0 the extension of CMG is largest in the z-direction, and hence the partial sum over the $ Z_{i,j}^2$ is largest. 
Going back in time the partial sums  are steadily decreasing. Already at -5~Myr the partial sum in $Y$ reached its minimum, strongly increases afterwards and dominates the value of $R$.  We rate this as a kind of "virtual expansion" caused by the poor knowledge of radial velocities (see also the scatter in the middle panel of Fig.~\ref{regression}).

If we neglect the $Y$-component, and take only the 2-D differences the minimum for the 2-D $R$ is reached at -7.2~Myr for all 123 stars. We show the back-tracing results in the $X,Z$-plane in in Fig.~\ref{XZback}.
From left to right we show the distributions of the 123 stars at present, 9\,Myr ago and 18\,Myr ago. The positions of the stars 18\,Myr ago are shifted by 50\,pc in the positive $Z$-direction to avoid overlap with the situation at -9~Myr.
Already at $-9$\,Myr and even more at $-18$\,Myr we notice 11 outliers. If we exclude the latter from the calculation of the radius $R$, we find a minimum of the 3-D $ R$  at -7.0~Myr, and of the 2-D $ R$  at -9.2~Myr.

\begin{figure}[htb!]
  \centering
  \plotone{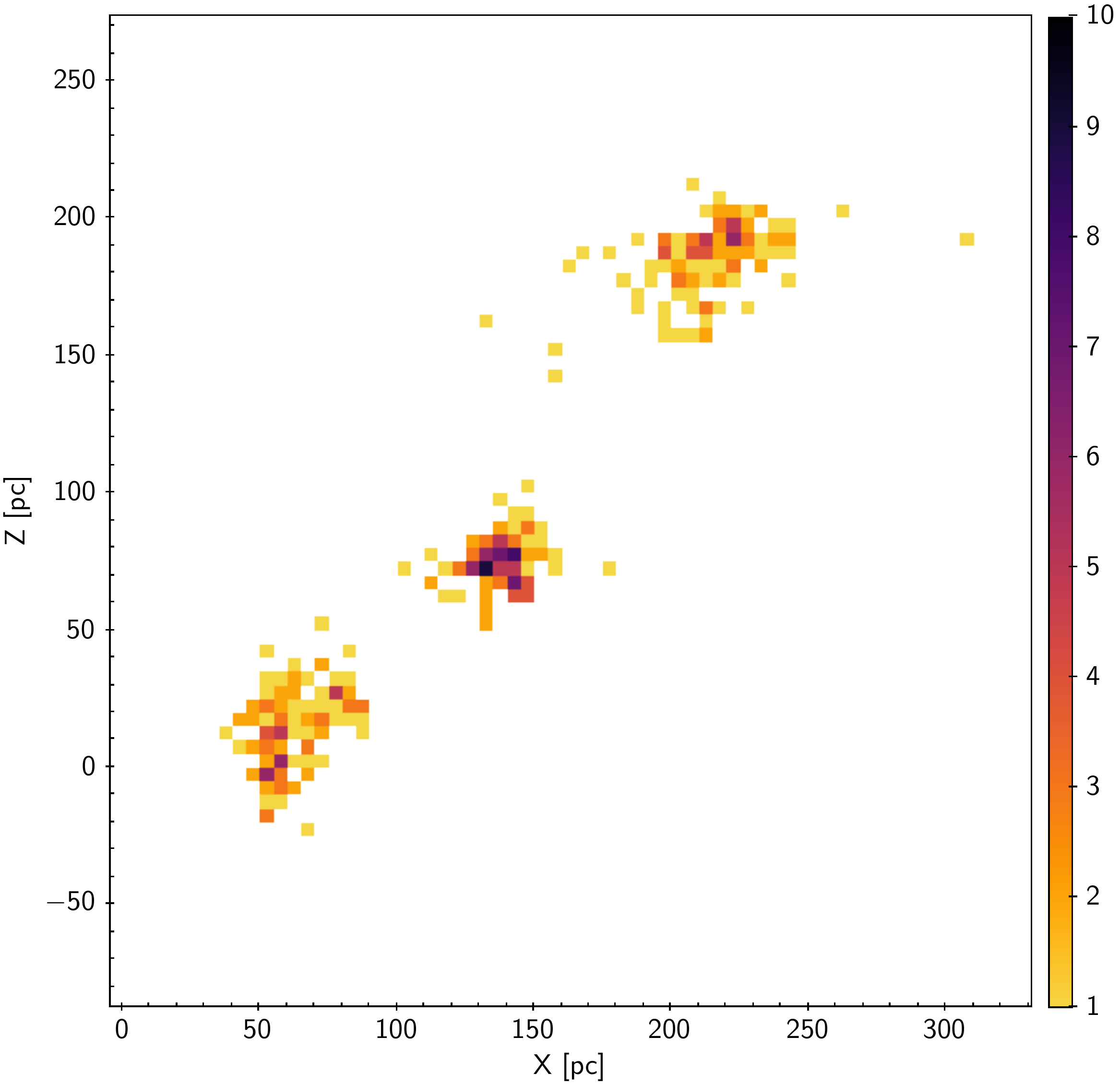}
  \caption{Density distribution of 123 CMG stars having accurately measured radial velocities (see text). The most left distribution shows the configuration at present, the central one 9\,Myr ago and the right one 18\,Myr ago. The latter is shifted by 50\,pc in the positive $Z$-direction to avoid overlap with the configuration at $-9$\,Myr.}
  \label{XZback}
\end{figure} 

We conclude that the CMG as a whole is expanding since at least 9 to 10\,Myr (group centers), and the sub-groups A0 to C, which we observe today, were parts of a single structure at that time. The clumps are probably the results of slightly different initial space velocities. We cannot exclude that expansion already started more than 10\,Myr ago as back-tracing is hampered by the poor knowledge of radial velocities ($2\,\rm km~s^{-1}$ at 10 Myr make an offset of 20 pc). Although 70\% of the stars in CMG were estimated to be younger than 10 Myr, the onset of star formation could have started earlier, and a later event triggered expansion. 

\citet{2018MNRAS.476..381W} explicitly stated that they found no indication of expansion in their careful analysis of the kinematic history of LCC on the basis of \Gaia\ DR1, except that they got evidence for expansion along the $Y$ axis with a significance of almost 2$\sigma$ for their LCC sample. Also,
\citet{1992A&A...262..258D} did not detect an expansion in the region of Crux, but from a completely different approach.} He examined the distribution of neutral Hydrogen gas (H1) connected to the Sco-Cen association. Whereas the column density of H1 is low over most of the area of CGRS, he found an H1 loop (shell) at $(l,b) = (285\deg,+18\deg)$, and determined its center to be at $(l,b) = (300\deg,+8\deg)$ from the curvature of the loop.  He estimated the mass of the molecular cloud in the loop to be $(1.0\pm 0.5)\times10^{5}$ M$_\odot$. Remarkably, the center of the loop is situated in the middle between our groups B and C, and the loop itself marks the northern border of group C at this Galactic longitude. From his H1-21cm observations he found no evidence for this loop of being part of an expanding shell.

Contrary to \citet{1992A&A...262..258D}, \citet{2009AJ....137.3922O}, who also made an analysis of the motion of the H1 shell in Crux, found that the shell had been expanding. Between 13 and 9 Myr before present the shell increased its radius from
15 to almost 30~pc, but then met together with the UCL shell. In this case this collision could have triggered the star formation of CMG which had its maximum about 9 Myr ago.
 
{
\section{Summary and discussion}\label{conclusion}
The data from the second \Gaia\ data release are an invaluable tool to study the kinematics of stellar ensembles in the Milky Way. Moreover, due to its excellent photometric data, \Gaia\ enables accurate determinations of stellar masses and ages from color-absolute magnitude diagram. However, it becomes apparent that that the wealth of new observations of low-mass stars in the \Gaia\ passbands requires new efforts to construct adequate theoretical isochrones.
We studied in this paper the large complex in Lower Centaurus Crux where a few million years ago star formation has taken place. We found some 2800 stars in a cone around \mbox{$(\alpha,\delta)$ =  (186.5\degr, -60.5\degr)} with a radius of 20\,degrees, which we could rate as younger than 40\,Myr based on the location in the ($M_G$, $G-G_{\rm RP}$) CMD. Among these, 1844 stars share tangential motion compatible with a common 3-D space motion. Their distances from the Sun range from 102 to 135 pc. The total mass of the stellar and sub-stellar objects is about 700 M$_\odot$. On top of this common motion the whole group is expanding, and expansion started about 9\,Myr ago.  Also 9\,Myr ago the peak in the star formation rate occurred with  $8\times10^{-5}\,\rm M_\odot yr^{-1}$. 

So, the event that was responsible for the beginning of expansion could also have triggered a higher star formation rate. Or, did a collision between UCL and LCC molecular shells occur 9\,Myr ago, as \citet{2009AJ....137.3922O} claim, and did such a collision trigger star formation. }

\EuS{The ages we determined for the stars in CMG are consistent with the age of the LCC according to \citet{2012AJ....144....8S}. 
Moreover, although \citet{2016MNRAS.461..794P} determined  an age of 17\,Myr for their LCC sample, their average 
age of the subset in common with CMG is identical with ours within 20\%, proving that the mean age 
of CMG is quite reliable.
Similarly, }
we find that the LCC {association} is much younger than the 18\,Myr found by \citet{Mamaj02}. This may have the following reason. Their age determination was based upon main-sequence turnoff ages for Hipparcos B-type members. The vast majority of stars in our sample have masses less than 0.3 M$_\odot$, and the ages of these pre-main-sequence stars and brown dwarfs are determined by comparing the three \Gaia\ DR2 photometric bands (absolute magnitudes) with the theoretical isochrones of CIFIST, and we rely on these age marks.  As the ages from \citet{Mamaj02} are based upon high-mass stars, there may be some inconsistency on the models underlying both methods. 

{
 Given the youth of the group, the present-day mass function should not differ too much from the initial mass function of the complex. We find that a log-normal mass function with mean mass $m_c=0.22$\,M$_\odot$ and a standard deviation $\sigma=0.64$ fits the observations quite well between 0.05 and 1.0M$_\odot$. We observe, however, a steeper decline in the brown dwarf regime at masses lower than 0.05, which cannot be attributed to incompleteness of our sample. In total, CMG contains 214 brown dwarfs. There is also an excess at about 1.4 M$_\odot$ which may have been caused by an imperfect separation between young and old stars near this mass. Our log-normal mass function differs from the canonical Chabrier mass function \citep{Chabr03a} essentially by its broader standard deviation (0.64 vs. 0.55). A Chabrier type mass function was also found by \citet{2013MNRAS.431.3222L} when he studied the star formation in the Upper Scorpius part of Sco-Cen.
 While the slope of the Kroupa mass function \citep{Kroup02} describes the high-mass end quite well, it fails at low-mass stars and brown dwarfs. First, the observed maximum of our MF is at higher masses (0.2 observed vs. 0.1 M$_\odot$ in Kroupa), and our slope in the brown dwarf regime is much steeper.
The latter is also valid for a comparison of our mass function in LCC with the mass functions in the $\sigma$\,Orionis cluster \citep{Bejar11,PenaR12}, and the very young embedded open cluster RCW 38
\citep{2017MNRAS.471.3699M}.} 

 {
With members' ages ranging from 5 to 20\,Myr, we detected no less than 170~circumstellar disks showing a wide variety of evolution, from full disks to debris disks. Among these, 34 are associated with brown dwarfs. The BD disks are in its majority full/evolved, and 14\% of the brown dwarfs in our sample carry these.}
 
{
Summing it all up, LCC is a rich {association} very close to the Sun, and this makes it an ideal laboratory for studying very young stars and very young brown dwarfs and their circumstellar disks.
The total stellar mass of our sample in LCC is about 700\,M$_\odot$, and 80\% of its stars are located d between 102 and 135 pc from the Sun with a median distance of 114.5 pc. 
According to the ages of our objects determined from the CIFIST isochrones star formation started about 20 Myr before present and reached its maximum 9 Myr ago. This is equal to the expansion age of our CMG.
Hence the event that was responsible for this expansion, may also have been an additional trigger for star formation}.

\BG{Our findings and our comparison with previous studies of young stars in the west part of Sco-Cen, show how the much deeper and more accurate astrometry of {\em Gaia} will shed light on moving groups, and will require a similarly more structured classification of the known moving groups.}

\acknowledgments

{We thank the anonymous referee for her/his comments that helped improving the manuscript.}
We are grateful to Ulrich Bastian from ZAH for very helpful discussions on the astrometric quality of \Gaia\ DR2 and on the procedures needed to obtain and astrometrically clean sample of stars.

This study was supported by Sonderforschungsbereich SFB 881 ``The Milky Way
System" (subprojects B5 and B7) of the German Research Foundation (DFG). 
This research has made use of the SIMBAD database and of the VizieR catalogue access tool, operated at CDS, Strasbourg, France.

This work has made use of data from the European Space Agency (ESA)
mission \Gaia\ (\url{https://www.cosmos.esa.int/gaia}), processed by
the \Gaia\ Data Processing and Analysis Consortium (DPAC,
\url{https://www.cosmos.esa.int/web/gaia/dpac/consortium}). Funding
for the DPAC has been provided by national institutions, in particular
the institutions participating in the \Gaia\ Multilateral Agreement.

This publication makes use of VOSA, developed under the Spanish Virtual Observatory project supported from the Spanish MICINN through grant AyA2011-24052.

This publication makes use of data products from the Wide-field Infrared Survey Explorer, which is a joint project of the University of California, Los Angeles, and the Jet Propulsion Laboratory/California Institute of Technology, funded by the National Aeronautics and Space Administration.

\facilities{Gaia, CDS, WISE} 

\software{Astropy \citep{2013A&A...558A..33A}}

\bibliography{bibiography}

\end{document}